%% file: main.tex
\documentclass[a4paper,11pt]{article}
\usepackage{jinstpub} 

\usepackage[utf8]{inputenc}
\usepackage{graphicx}
\usepackage{siunitx}
\usepackage{url}
\newcommand{\figref}[1]{\figurename~\ref{#1}}

\title{Radio emission from airplanes as observed with RNO-G}

\collaboration{\includegraphics[height=20mm]{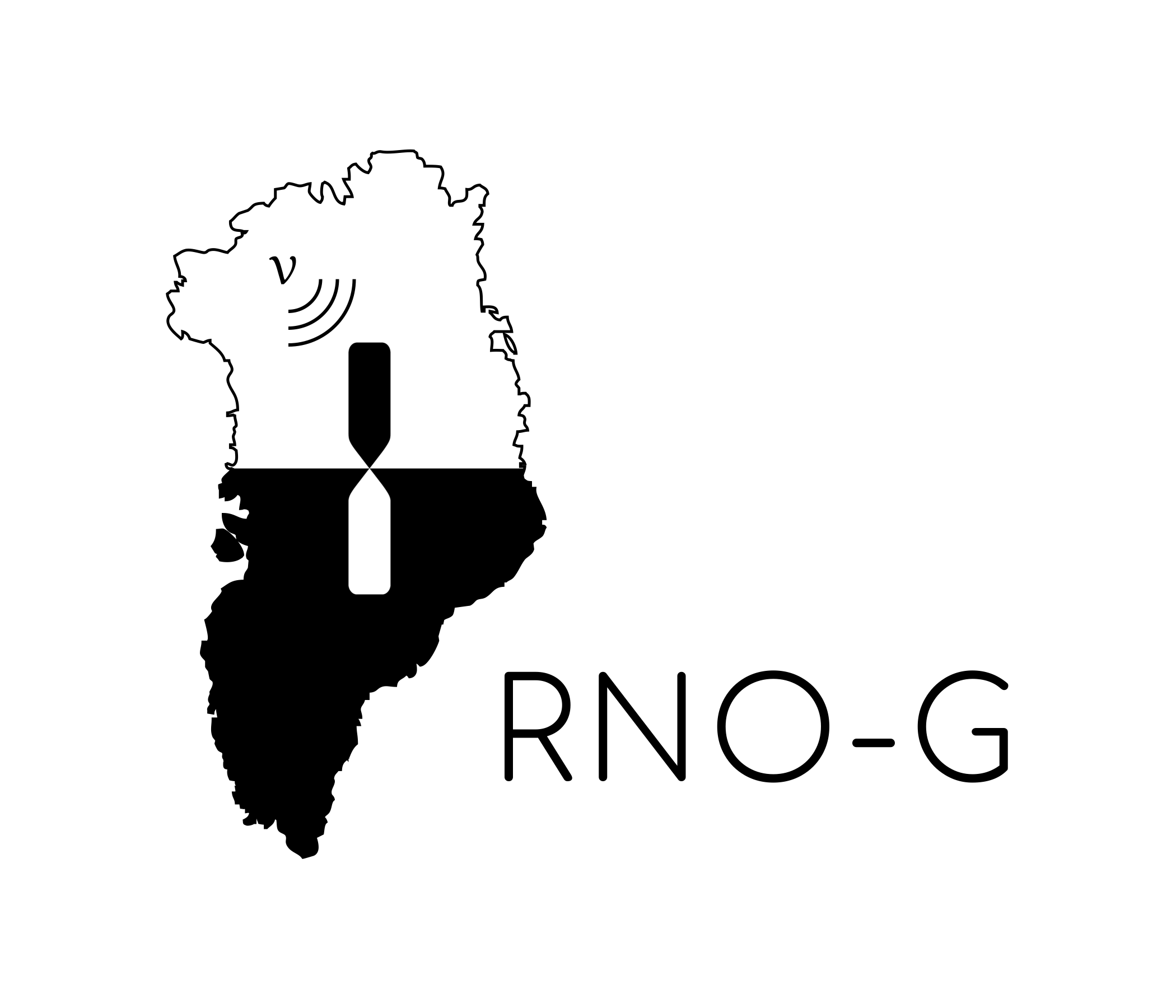}\\[6pt]
RNO-G Collaboration}

\input{authorlist}

\emailAdd{authors@rno-g.org}
\emailAdd{steffen.hallmann@desy.de}

\abstract{This paper describes how intentional and unintentional radio emission from airplanes is recorded with the Radio Neutrino Observatory Greenland (RNO-G). We characterize the received signals and define a procedure to extract a clean set of impulsive signals. These signals are highly suitable for instrument calibration, also for future experiments. A set of signals is used to probe the timing precision of RNO-G in-situ, which is found to match expectations. We also discuss the impact of these signals on the ability to detect neutrinos with RNO-G. }


\arxivnumber{1234.56789} 

\begin{document}
\maketitle
\flushbottom

\section{Introduction}

Airplanes are known to emit impulsive radio signals that are measured with experiments seeking to detect radio emission from astrophysical sources or their messengers, e.g.~\cite{PierreAuger:2015aqe,Charrier:2018fle,Monroe:2019zkp,2020PASA...37...39T,2023ApJS..265...62R,2024A&A...681A..71G,Sokolowski:2021rao,Zhang2004,Ducharme_Pober_2025,rs13040810}. While they have the potential to spoil the main astrophysical purpose of the telescopes, they can also be used to calibrate the instrument with respect to timing and absolute pointing. In particular for large arrays of instruments in inaccessible areas, these signals provide calibration opportunities that are not easily matched. 
While discussing the implications of airplane signals for neutrino searches, this paper will focus on the calibration needs and opportunities of large arrays for the detection of the radio emission stemming from neutrinos above $10^{16}$ eV like the Radio Neutrino Observatory Greenland (RNO-G) on the top of the ice sheet of Greenland. 

\begin{figure}
    \centering
    \includegraphics[width=.48\textwidth,trim={8cm 0 5.9cm 0},clip]{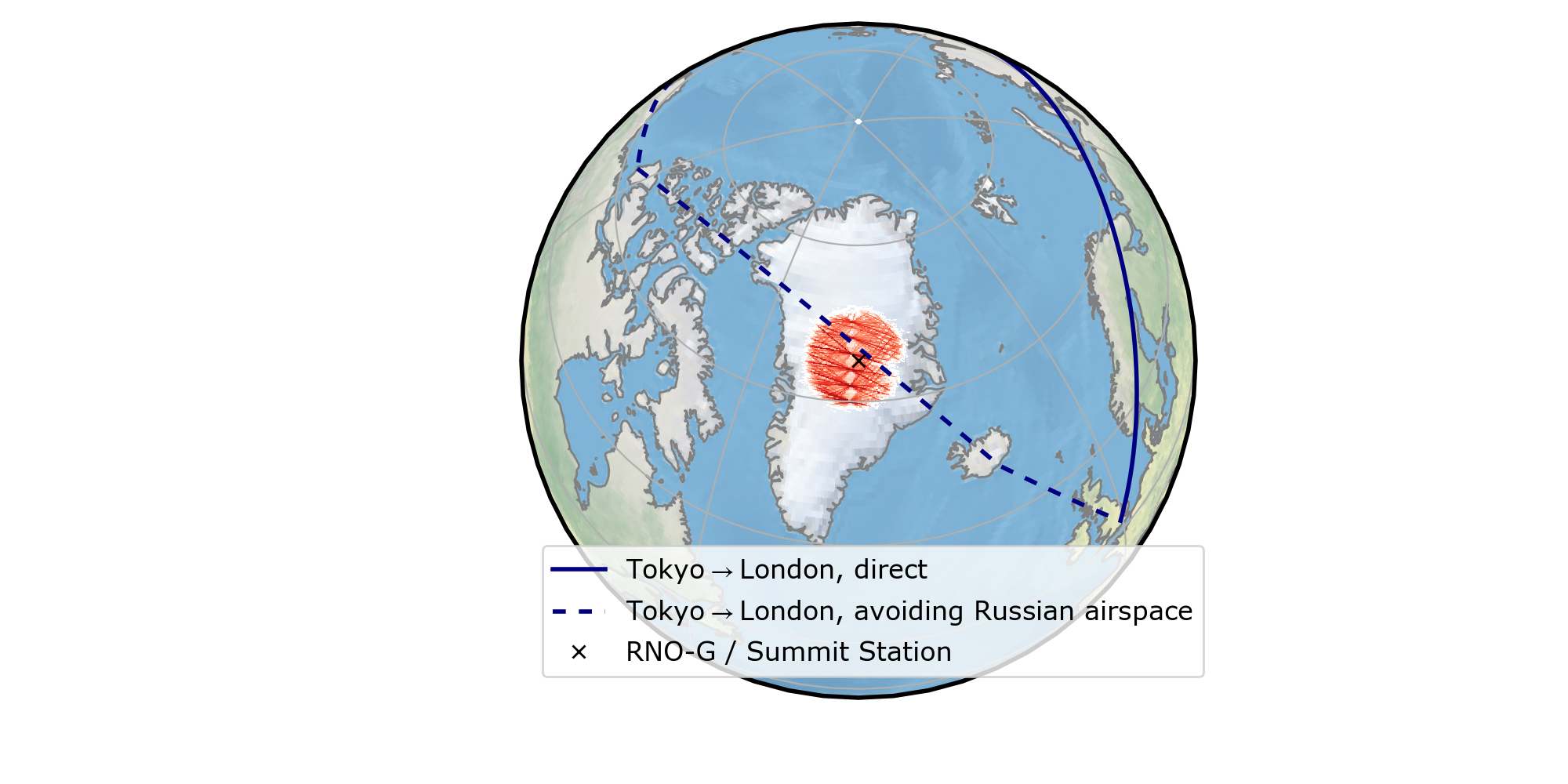}
    \includegraphics[width=.5\textwidth,trim={1cm 0cm 3cm 0},clip]{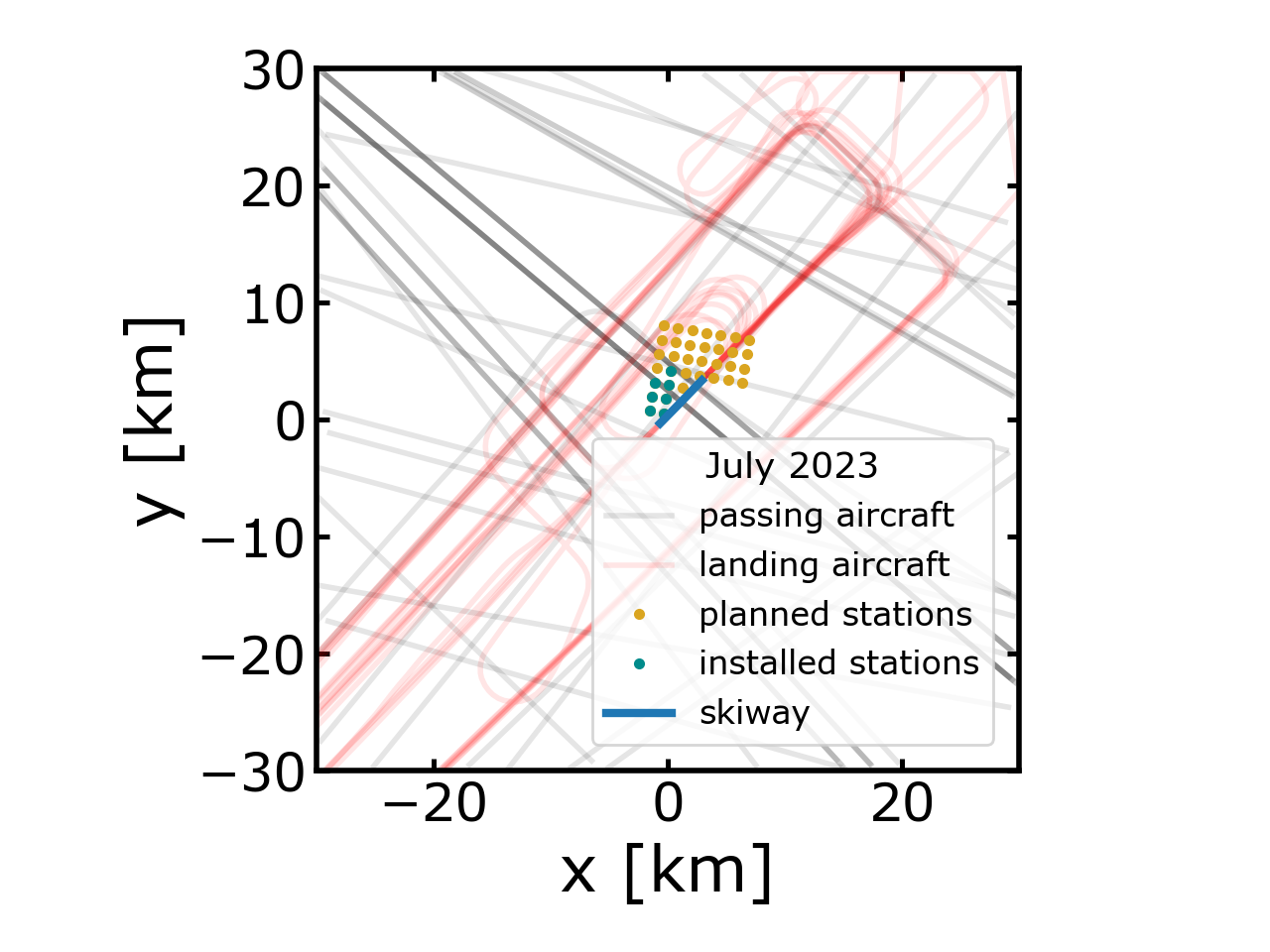}  
    \caption{\textit{Left}: Location of Summit Station in Greenland with approximate flight paths between Japan and Europe. The colormap (white to red) indicates the density of aircraft locations seen by the ADS-B receiver up to $\sim$350\,km around Summit Station. \textit{Right}: Zoom on RNO-G array (every point represents one station as shown in \figref{fig:rno-g-station}) together with all recorded flight paths from July 2023. The airplanes landing on the skiway at Summit Station are highlighted in red. }
\label{fig:greenland_map_with_adsb_positions}
\end{figure}

\subsection{RNO-G}
\begin{figure}
    \centering
    \includegraphics[width=0.4\linewidth]{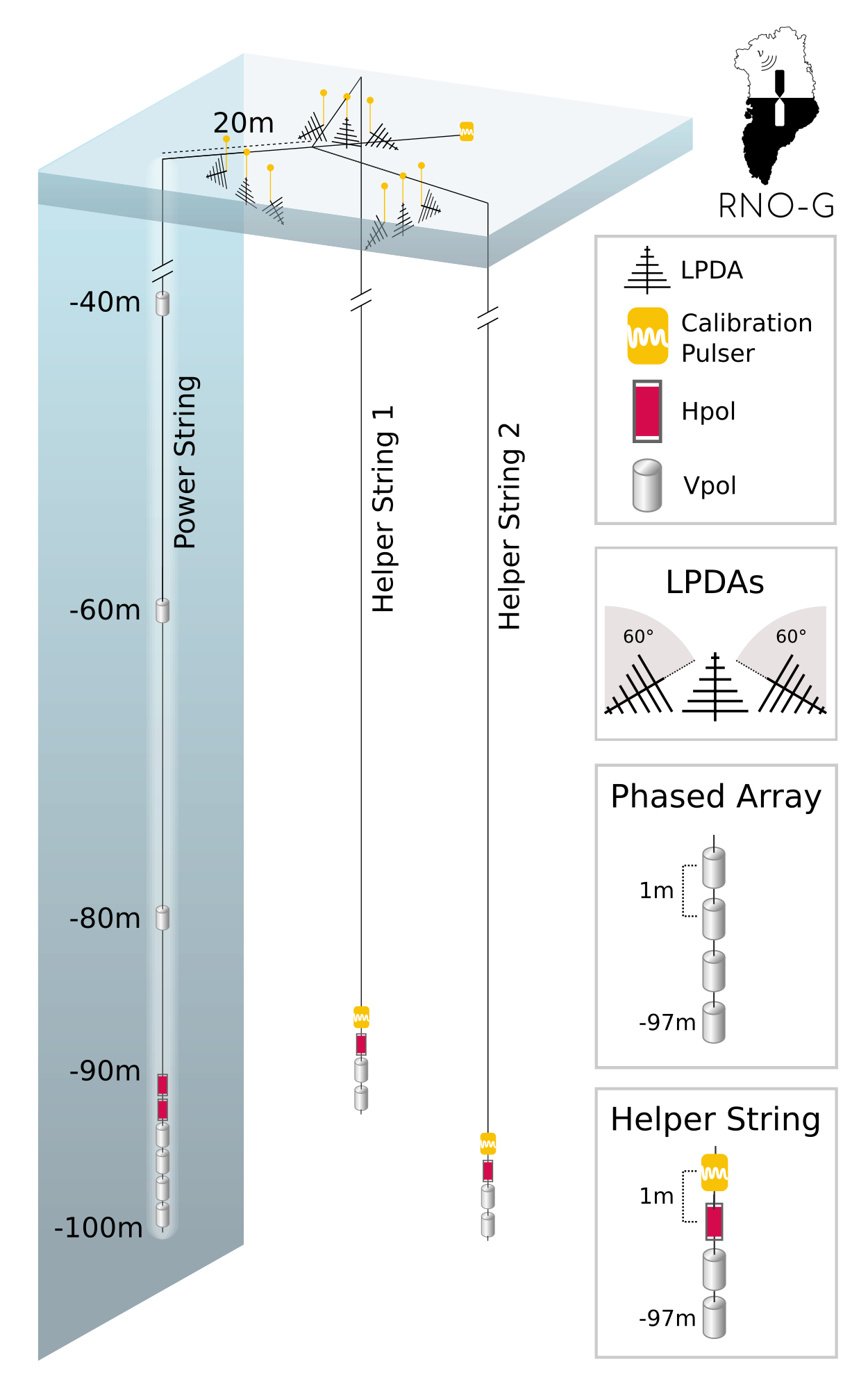}
    \caption{Schematic view of an \hbox{RNO-G} station. Each station has 24 antennas, either LPDA, Hpol or Vpol (see text for details) and three in-situ pulsers for calibration purposes. LPDAs are deployed in shallow trenches while Hpol and Vpol antennas are lowered into deep dry boreholes. Each point in the array map of \autoref{fig:greenland_map_with_adsb_positions} corresponds to one of these stations.}
    \label{fig:rno-g-station}
\end{figure}

RNO-G utilizes the radio transparent properties of polar ice \cite{Aguilar:2022ijn} that allow to detect the faint radio pulses that result from the deep-inelastic interactions of astrophysical neutrinos with nuclei near the detector \cite{RNO-G:2020rmc}. The radio signals are emitted on a cone in the forward direction and are mapped with an array of antennas that are embedded in the ice. RNO-G is planned to consist of 35 stations on a rectangular grid with a spacing of \SI{1.25}{km}, installed near Summit Station in Greenland (\autoref{fig:greenland_map_with_adsb_positions}). Stations (see \autoref{fig:rno-g-station}) are installed during the arctic summer, with a total of 8 stations operational 2024. Each station consists of 24 antennas located at different depths down to \SI{100}{m}. The signals from these antennas are recorded locally per station after a trigger and are then forwarded over a wireless network to a central processing facility \cite{RNO-G:2024esr}. Three different types of antennas are used in RNO-G that cover frequencies from just below \SI{100}{MHz} to \SI{700}{MHz}. Determined by the size of the boreholes, custom-made Vpol and Hpol antennas (sensitive predominantly to the vertical and horizontal signal polarization, respectively) are used in the ice to map out the neutrino signal. Just below the surface, large commercial logarithmic-periodic dipole antennas (LPDAs) are installed. The LPDAs are installed at different angles to be sensitive both to the upward going neutrino signals, as well as downward going air shower signals and human-made backgrounds. 

RNO-G does not continuously record data, but saves a millisecond-long snapshot of raw data after a trigger has been issued based on signal characteristics. RNO-G has a number of different triggers. The current main trigger uses the compact group of four Vpol antennas (\emph{phased array}) at the bottom of one of the holes in each station. This trigger is the most sensitive to small impulsive neutrino signals and is issued once a threshold is exceeded in the combined signal in a small coincidence time-window. In addition, triggers are issued if impulsive signals are measured in coincidence of upward-facing LPDAs or downward-facing LPDAs. The LPDA-based triggers run at a higher signal threshold, because they are allocated only 10\% of the trigger budget.  However, they arrive at the data acquisition first, so de-facto take priority over the Vpol-based triggers, unless the maximum read-out rate is already reached.  

As a result of its remote location, RNO-G relies on solar power. This reduces data taking to the months when it is light at the RNO-G latitudes, between March and October. While wind-turbines have been tested at some stations, the additional up-time from wind-power is still limited, but it is expected that this can be improved in the future. 
More technical details about RNO-G can be found in \cite{RNO-G:2020rmc} and the recently published performance results \cite{RNO-G:2024esr}.

\subsection{Calibration needs}
The holes for the RNO-G antennas are drilled with a large mechanical drill. While all antennas have nominal positions, slight tilts of the boreholes, as well as misalignment of the installation, lead to positioning uncertainties \cite{RNO-G:2024esr}. Since the antennas are installed deep in the ice, they cannot be surveyed using GPS after installation. Similarly, keeping track of snow accumulation and ice movements through surveying requires a significant effort. In order to reduce this effort, all stations are equipped with local calibration pulsers \cite{Oberla:2023Q8}. For a \emph{global calibration} of the antenna positions in each separate station with respect to the GPS reference, however, the system is underconstrained, because of the correlation in position offset and ice properties, in particular the index of refraction profile as a function of depth \cite{Beise:2022stx,Windischhofer:2024iK}. Also, cable delays measured in the laboratory change with lower temperature which requires a correction factor that comes with an uncertainty and thus leads to an additional systematic effect.  

Thus, well-located signals that illuminate (several) stations from afar are a relevant additional calibration source. Recently, RNO-G has shown that solar radio flares can be used for antenna position calibration \cite{Agarwal:2024tat}. However, these happen most likely during the current solar maximum and may not occur frequently enough to calibrate all future stations. Furthermore, the solar signals are only partly impulsive, which complicates finding matching signals across all antennas, and the Sun never rises fully overhead in Greenland, which limits the coverage in zenith angles that can be probed with this calibration approach. 
Alternatively, experiments have used balloons or drones that fly a predefined calibration pattern over the array, e.g.~\cite{Nelles:2015gca,Prohira:2017sal,TAROGE:2022soh}. With these dedicated drones, high-amplitude impulsive signals are possible. However, drone calibration campaigns are expensive, take significant effort, and are therefore typically only done once.

Consequently, airplanes flying over the array may be a very suitable calibration source, offering a rich source at a well-known location in the far field at a broad range of angles with high-amplitude impulsive signals. In particular, the location in Greenland is interesting, as several airline routes connecting Europe with Asia and North America cross the country with a direct line of sight of RNO-G, as shown in \autoref{fig:greenland_map_with_adsb_positions}. Before being able to define a procedure to calibrate the antenna positions and study the obtainable accuracy, it is necessary to characterize the airplane signals and select a set for the purpose of calibration. This selection procedure is the subject of this article. 

The suitability of the event set for calibration will be tested in two ways. We will cross-check the positions of the LPDAs obtained by and differential GPS (dGPS) survey and by measuring the timing accuracy of the RNO-G stations in situ, with the latter being a performance metric rather than a calibration.

\subsection{Aircraft identification and their signals}

During flight, most airplanes broadcast information about their flight trajectory using Automatic Dependent Surveillance–Broadcast (ADS-B) messages transmitted at \SI{1090}{MHz}. These messages contain the aircraft's unique identifier and name, as well as the in-air position, orientation, speed, and vertical ascend rates. The ADS-B messages are recorded with dedicated equipment at Summit Station, ensuring that everything in the field of view is recorded. Publicly available databases of ADS-B messages typically do not cover Greenland. 

The position of each aircraft is determined via GPS or another GNSS and should hence be accurate to a precision of a few meters. The ADS-B receiver at Summit Station logs the time at which messages are received. The time and position data can hence be used to obtain the time-dependent trajectory for calibration. However, for some flights the information transmitted via ADS-B is incomplete. 

\begin{figure}
    \begin{minipage}{0.49\textwidth}
    \centering
    \includegraphics[width=\linewidth]{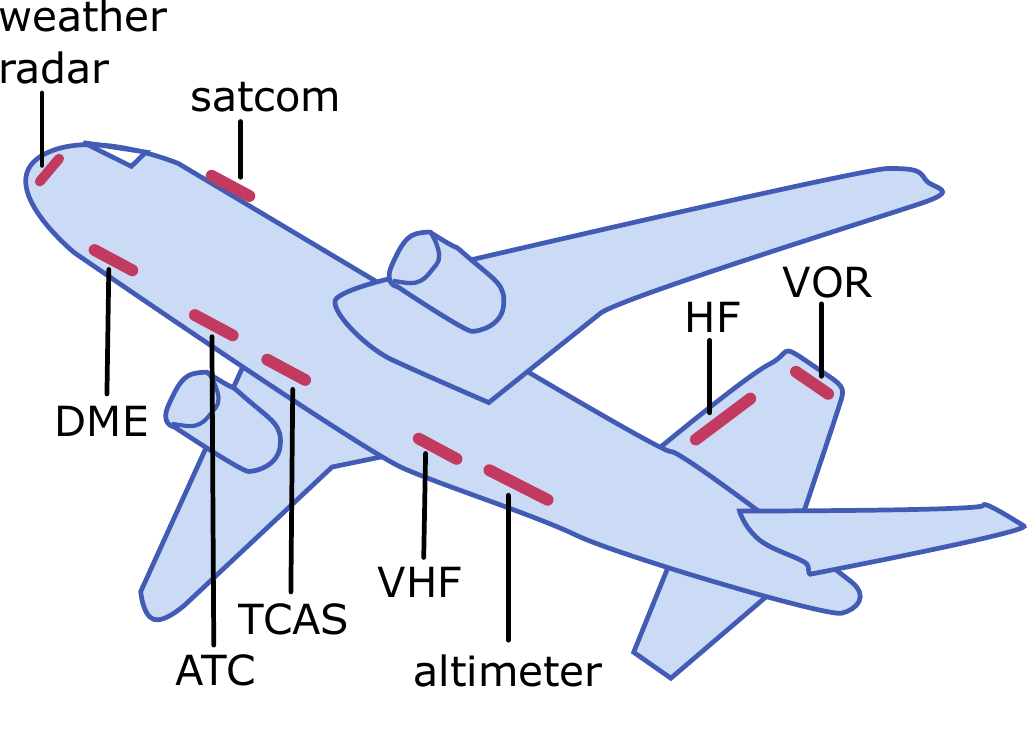}
    \end{minipage}    
    \begin{minipage}{0.49\textwidth}
    \caption{Key transmitting antennas as potential RF sources on a commercial aircraft, see e.g.~\cite{Airplane_forum} for a detailed explanation of all acronyms. In case of redundancy only one position is indicated per probe type. The turbines are also expected to generate measurable signals \cite{LOFAR_Airplane}.}
    \label{fig:airplane_schematics}
    \end{minipage}    
\end{figure}

The ADS-B messages are not the only signals received from airplanes. Airplanes emit both unintentional emission (RFI) from electrical devices such as the turbines, and intentional radio signals, such as from radar altimeters. The typical set of transmitting antennas installed on airplanes is summarized in \autoref{fig:airplane_schematics}. Depending on the type of aircraft, many different emission points are possible, such as high-frequency antennas in the tail for plane-to-plane communication, radar altimeters at the bottom of the aircraft, or emergency locators at various positions on the rump of the aircraft. Even if their nominal transmission frequency is outside of the RNO-G band, they cannot be excluded as sources for RNO-G, as intermodulation or resonances can make signals detectable in the RNO-G band.

We will not attempt to locate and disentangle all different types of emission as dedicated radio telescopes are more suitable for that, e.g.~\cite{LOFAR_Airplane}.
Still, we will show that distinct signatures can be identified by RNO-G and that a clean set of impulsive signals can be selected for calibration purposes and will show how the timing accuracy of RNO-G can be probed using these data. Furthermore, we will discuss the potential impact of airplane signals on neutrino searches and discuss further steps.

\section{Airplanes recorded over RNO-G}
In this work, we use a period between August of 2021, right after the first RNO-G stations were installed, until the fall of 2024, before the stations turned off data recording for their winter mode.

\subsection{Flight tracker and information available}
\begin{figure}
    \centering
    \includegraphics[width=0.5\linewidth]{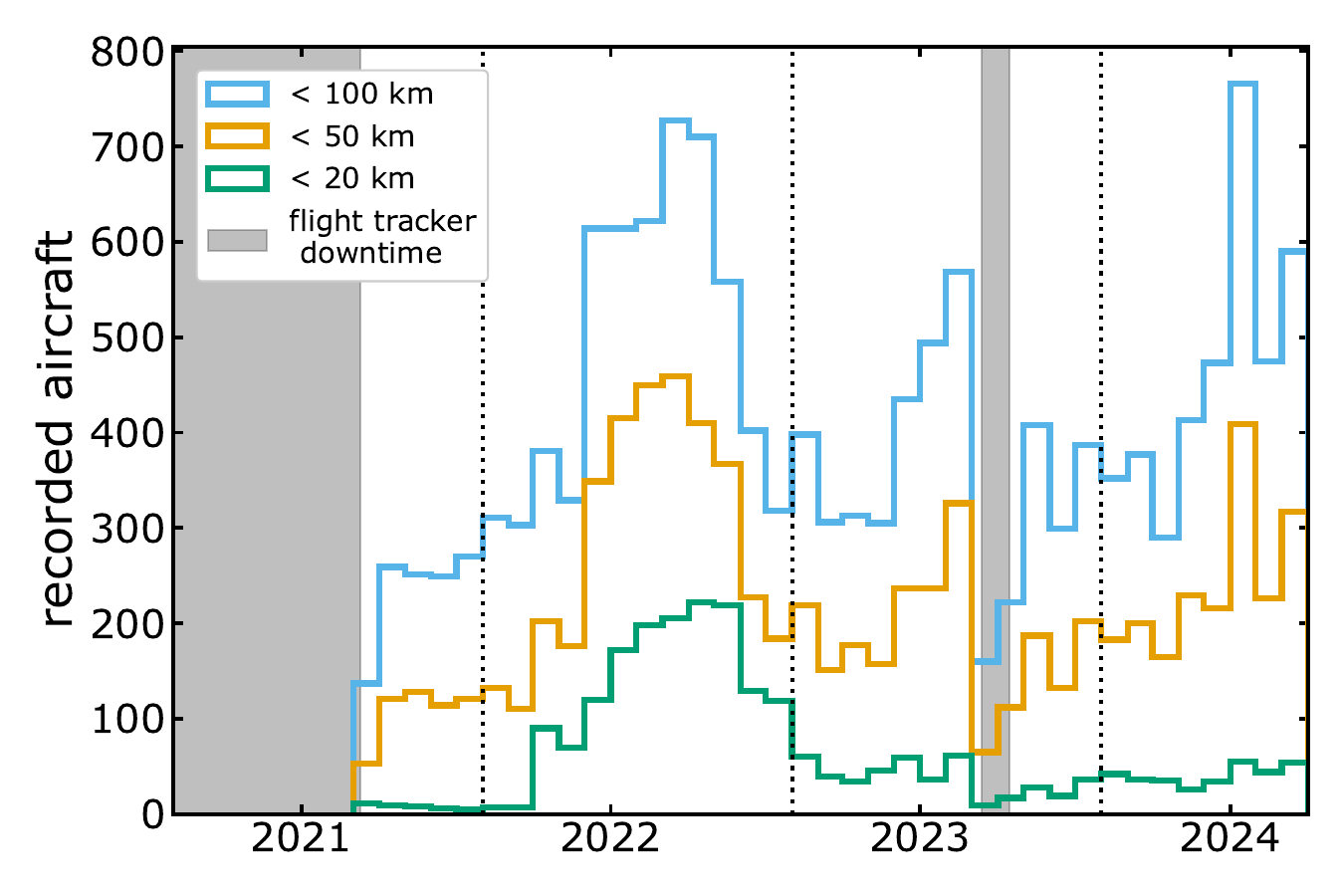}
    \hfill
    \includegraphics[width=0.49\linewidth]{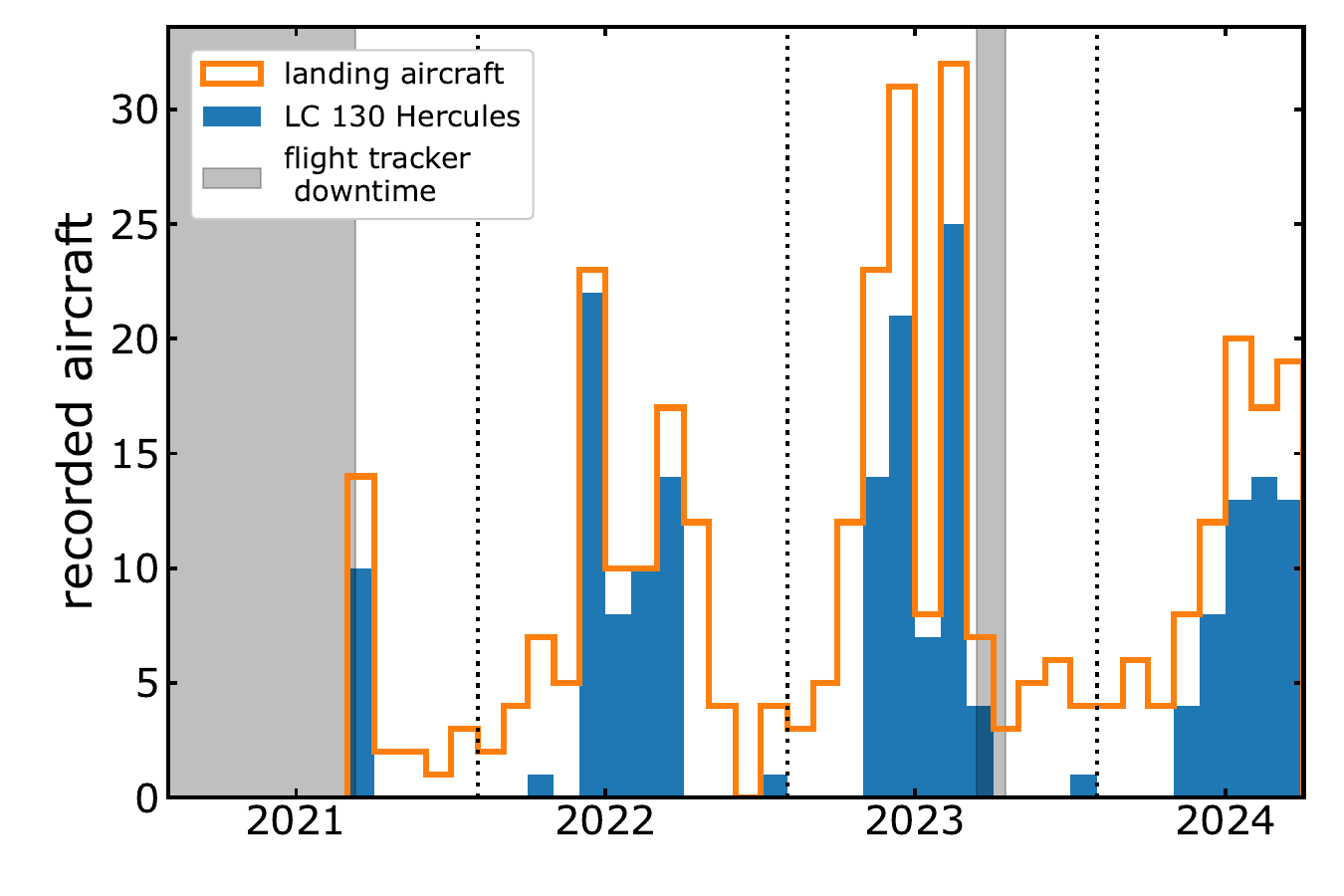}
    \caption{Statistics of passing flights recorded per month by the plane tracker at Summit Station. Periods of  missing data are marked with the gray bands. \textit{Left:} All airplanes broadcasting position information passing RNO-G at different distances as function of time. \textit{Right:} All airplanes landing and starting at the Summit Camp skiway. The fraction that is made up of the largest planes (LC130) is highlighted in blue.}
    \label{fig:aircraft_statistics}
\end{figure}

\autoref{fig:aircraft_statistics} provides an overview of the airplanes recorded by the ADS-B receiver at Summit Station, which was installed by members of the RNO-G collaboration. 
The flight-tracker had a continuous up-time, with the exception of two months in 2023, where an oversight led to an irrecoverable data loss.

For every flight, the ADS-B messages contain the time of the message (accuracy of \SI{0.1}{s}), the position (latitude, longitude, and altitude), the speed of the aircraft, as well as airplane identification information, such as an unique hexcode or the flight-number. Not all airplanes transmit valid entries for all fields.
Using publicly available databases on the web (such as \cite{AirplaneDB}), the hexcode and/or flight numbers can be matched to a specific aircraft, thereby providing information about for instance the number of engines or country of origin. 
From the received messages, the global coordinates are converted to local RNO-G coordinates to be used for later analysis. Also, initial cross-checks with a few bright airplanes found no global timing offset that needed to be taken into account. 

\subsection{Aircraft statistics}
Returning to \autoref{fig:aircraft_statistics}, it is clearly visible that the number of flights varies throughout the year due to local activity at Summit and Greenland as a whole, as well as due to changing global flight routes above the array. 
For the analysis in this publication, we used a total of 8505 flights that approached Summit Station to a distance of less than \SI{50}{km} between August of 2021 and September of 2024. Of those, 1251 uniquely different airplanes were identified, meaning that the same aircraft often flies a similar route multiple times. The most common flights are airplanes with two (or four) jet engines, corresponding to 7444 (829) flights in the data period. These are most often commercial flights from Europe to Japan as indicated in \autoref{fig:greenland_map_with_adsb_positions}. 

An important subset are the flights to and from Summit station, which account for less than 10\% of all flights recorded above RNO-G. This number should be comparable for all large research stations with a groomed runway, with precise flight numbers scaling with the size and activity of the station. For instance, while Summit Station remains open in winter, the large LC130 airplanes only land during the summer months. 

Military airplanes or private jets do not always transmit their position or identification. While position information was absent in only 0.5\% of all flights with recorded ADS-B messages, the fraction was significantly higher (28\%) for the LC130 flights approaching and landing at Summit Station. This means that signals from these planes are potentially visible in the data, but cannot always be used for station calibration.

Overall, the statistics of airplane signals that are in principle available for a calibration analysis is large. We note that due to the same routes, often similar angles are probed and that not all airplanes are visible in the RNO-G stations in the same way with the same signal quality. Therefore we discuss the observed signals, before selecting a subset for calibration purposes.

\section{Observed signals in individual stations}

From simply looking at strong signals observed in coincidence with airplanes, it is clear that not all signals are the same. We first illustrate the signal types by highlighting two particularly suitable flights and then discussing statistic metrics of recorded signals.

\subsection{Selected flights to highlight performance}

\begin{figure}
    \centering
\includegraphics[width=.49\textwidth,trim={0 2cm 0 0},clip]{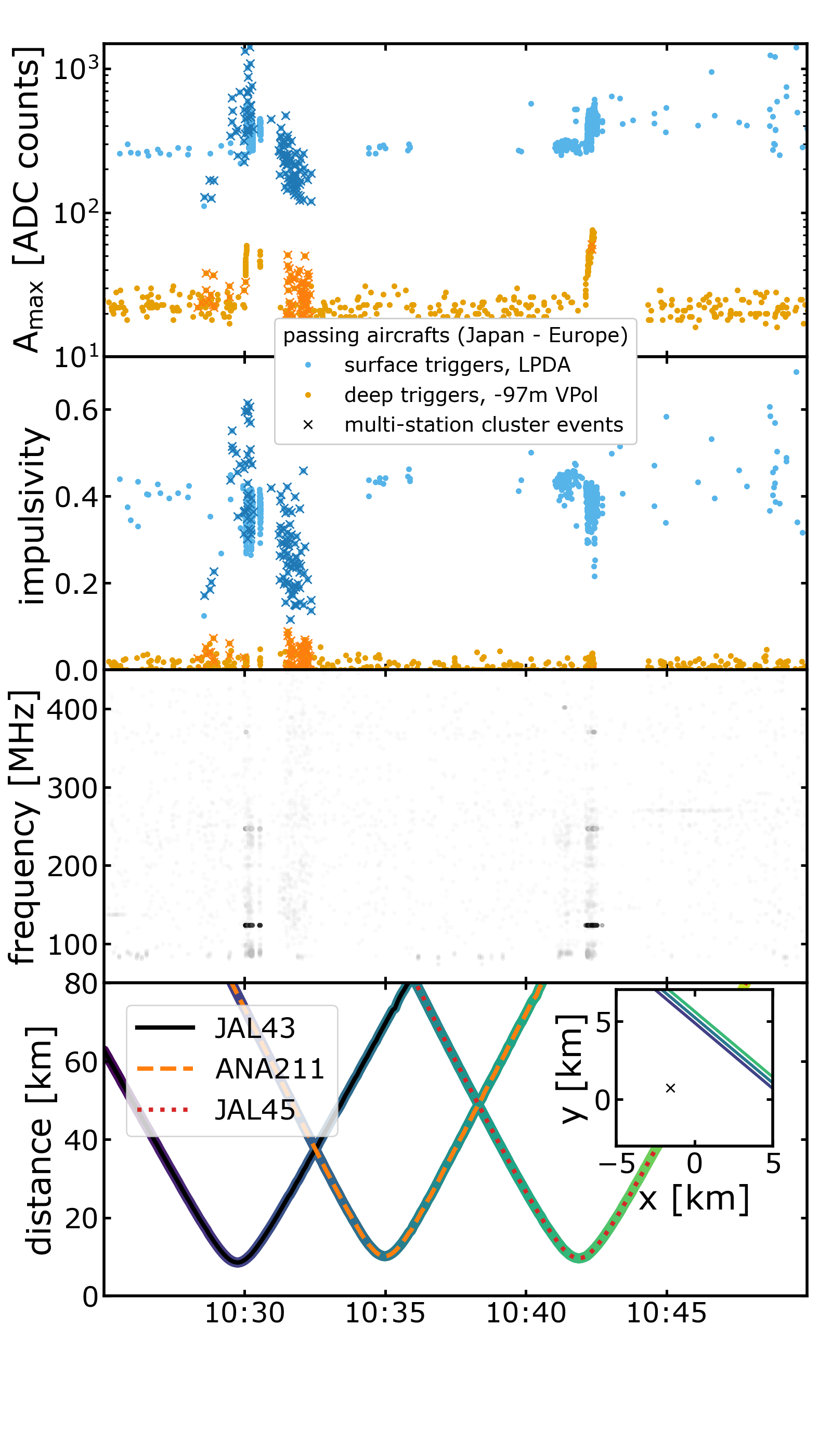}
\includegraphics[width=.49\textwidth,trim={0 2cm 0 0},clip]{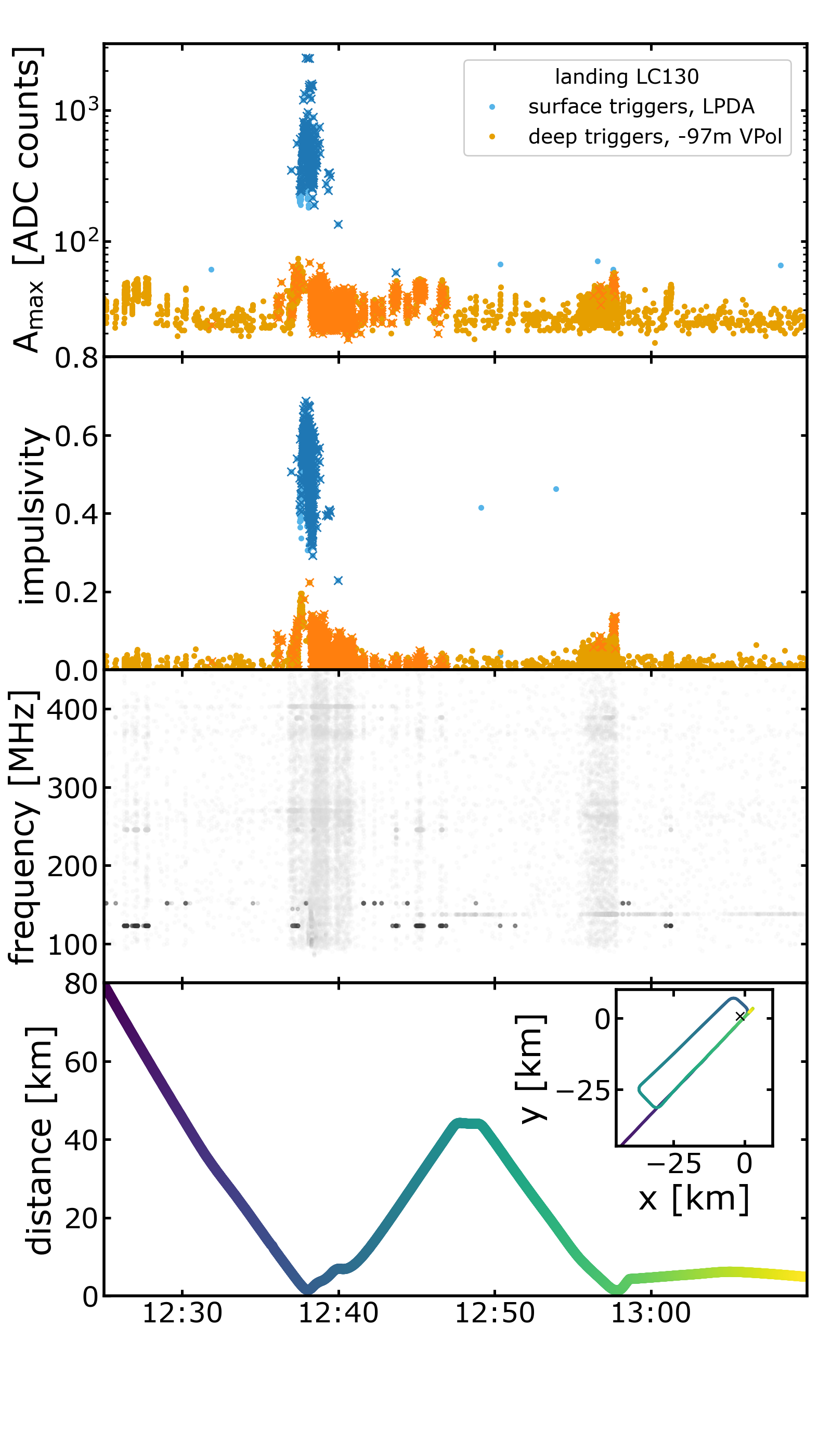}
\caption{Overview of the signals of two types of airplanes. \textit{Left:} Maximum signal amplitude, impulsivity ($I$), observed narrow-band frequency contamination ($CW$), and aircraft distance as function of time (UTC) for signals recorded in station 11 during the passage of three commercial airplanes on 08/30/2022. In the location displayed by the inset, the second and third track are shifted to the top right slightly -- in reality all three tracks follow the same path. Impulsivity is calculated after CW removal, see text for additional details. \textit{Right:} Same quantities for an LC130 cargo plane landing at Summit Station.}
     \label{fig:flight_event_properties}
\end{figure}

The first example consists of a series of three commercial airplanes flying almost back to back from Japan to Europe. The second example is an LC130 approaching Summit Station and completing a traffic circuit crossing the RNO-G array before landing on the skiway at station (see \autoref{fig:flight_event_properties}). 

In general, a distinction has to be made between impulsive signals and continuous wave signals (CW). Impulsive signals have a well-defined start-time and are very suitable for direction reconstruction and thereby position calibration. In contrast, continuous wave signals have no leading-edge and therefore cannot be uniquely aligned in time between stations, which makes them less suitable for calibration. Both types of signals have been found to be connected to airplanes.
To summarize the properties of the recorded events, we therefore use the maximum absolute amplitude, an impulsivity metric, $I$, and the strength of CW present in the waveform, $CW$. The impulsivity ranges from 0 to 1 and is calculated based on the excess in the Hilbert envelope of the waveform close to the highest amplitude pulse of the trace. It is defined as
\begin{equation}
    I = \mathrm{max}\left( 2\frac{\sum_i^N y_i}{y_{N}} - 1,\ 0 \right), \hspace{.5cm} y_i = \sum_{j=0}^i \left(h_\mathrm{sorted}\right)_{j},
\end{equation}
where $h_\mathrm{sorted}$ represents the values of the absolute Hilbert envelope re-arranged by increasing distance from the maximum value. The impulsivity $I$ (as first used in \cite{ANITA:2018vwl}) can thus be seen as a measure for the area above the diagonal of the histogrammed cumulative re-arranged Hilbert-envelope.
Strong $CW$ lines are identified and removed iteratively by fitting the dominant sine wave in the time domain outside the signal pulse region up to a 3$\sigma$-threshold above the average frequency spectrum after cleaning. The thus removed frequencies are shown in \autoref{fig:flight_event_properties} with a color-coding reaching from black (highest significance) to invisible (no significance).  

For the commercial airplanes strong CW signals at 123.45\,MHz are seen when the Japan Airlines (JAL) airplane are overhead, but are not visible for the All Nippon Airways (ANA) flight in-between. This frequency is used for air-to-air communication between airplanes emitted from the VHF antenna. High amplitude and impulsive events are also observed. The first flight is only partially contaminated with the CW, leading to many more multi-station coincident events compared to the second aircraft. The absolute number of triggers should be taken with a grain of salt, as the RNO-G trigger thresholds are changed to operate at a constant trigger rate, which means that very strong airplane signals temporarily "blind" the stations. Although the ANA aircraft does not cause multi-station coincident triggers, the density of triggered events around 10:34 UTC in \autoref{fig:flight_event_properties} does indicate that there is still signal recorded by the RNO-G stations.

For the landing LC130 aircraft (\autoref{fig:flight_event_properties}, right) more multi-station coincidences are seen from the rear of the aircraft compared to the front. In the second passage the aircraft is already too low for the signals to trigger the shallow LPDAs. During the approach, three clear CW emission frequencies can be identified --  a strong signal at 122.8\,MHz, which is the Common Traffic Advisory Frequency (CTAF)  typically used at airports without a tower, a weak signal at 137.8\,MHz typically associated with communication satellites, and a walkie-talkie frequency at 151.6\,MHz. A faint signal at 403\,MHz, associated with a weather balloon that is launched daily before noon and midnight is still discernible in the spectra.

\subsection{Characterizing airplane signals}

We can conclude from the flights shown above, that there is strong CW emission from aircraft communication (air-to-air and air-to-ground) which can trigger the RNO-G stations. CW signals do not cause multi-station coincidences within a small time window, because each station triggers on the long CW signal at a random time, when it first locally crosses a threshold. By saturating the trigger rate, these CW signals might even reduce the rate or multiplicity of multi-station coincidences. Consequently, they are not useful for calibration studies and have to be removed from the data-set.

Unless the trigger rate is saturated, strong and impulsive events trigger the shallow antennas if they are right above the stations, due to the trigger logic as discussed above. For large zenith angles the sensitivity of the LPDAs is reduced. During the approach and after the passage close to the station, the deep component is instead able to trigger thanks to the much lower threshold of the deep phased array trigger. However, the signal in the deep antennas is attenuated by the ice compared to the shallow antennas, such that the impulsivity and the maximum amplitudes are elevated but lower than for the LPDAs. Also, due to the trigger logic, the first arriving trigger will be read-out, meaning that in the current set-up the shallow trigger always supersedes a trigger from the deep antennas. For future stations, it is planned to add additional physical delays to the LPDAs also to better facilitate the recording of down-going cosmic ray signals. Such a set-up will allow for a more comprehensive study of signal attenuation and relative amplitude calibration of all channels at the same time. 

\begin{figure}
        \centering
        \includegraphics[width=\columnwidth]{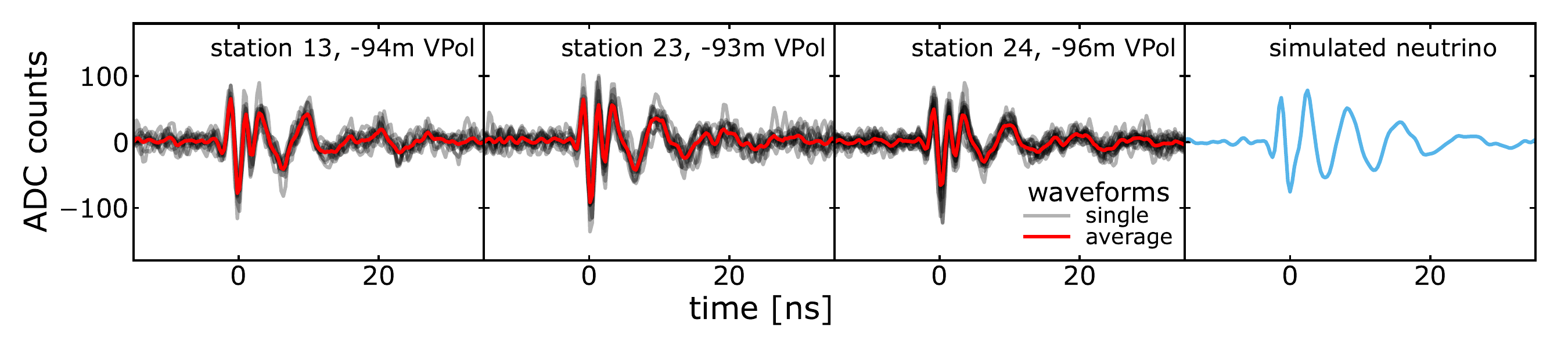}
        \caption{Highly impulsive events for a passing aircraft recorded with a Vpol channel and the same depth in different stations. The shapes of the waveforms are highly correlated with each other and between stations. The last panel shows a waveform of a simulated neutrino observed with a Vpol antenna for comparison.}
    \label{fig:similar_events_different_stations}
\end{figure}

\subsection{Event classes for impulsive events}

A general observation is that while planes pass by, similar signals are repeatedly emitted by the same emitter. Identical waveforms are not only recorded across multiple stations, but subsequent events are very similar in shape, as shown in \autoref{fig:similar_events_different_stations}. This repeatability provides the opportunity to generate averaged waveforms for flights for which there are sufficiently many triggered events of the same type. A set of simulated neutrinos on average showed lower correlation with the averaged waveform compared to individual aircraft events with each other. However, a simple impulsivity metric alone will not be able to separate signals from background.

\begin{figure}
    \centering
    \includegraphics[width=.52\linewidth]{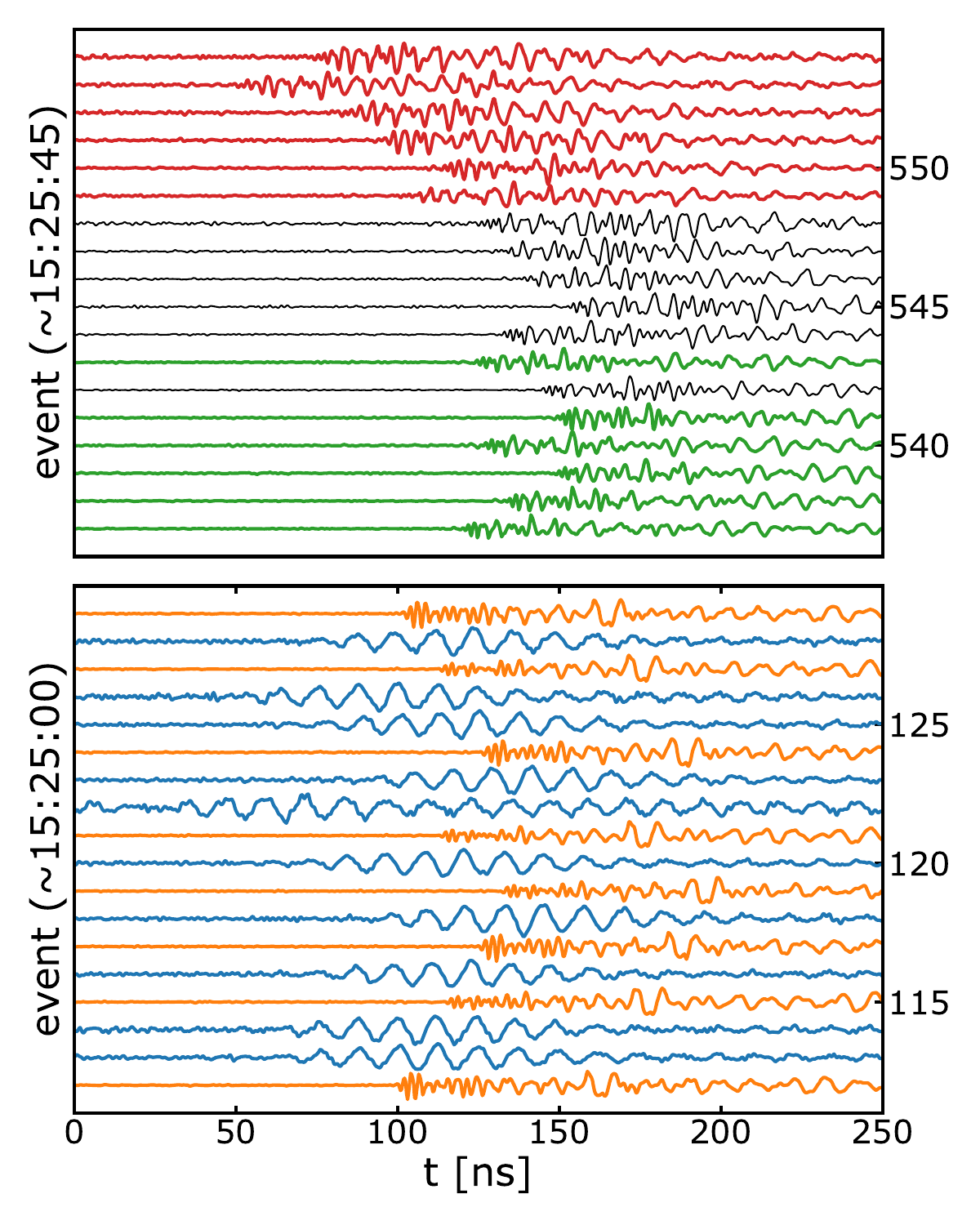}
    \hfill
    \includegraphics[trim=1.5cm 0 0 0, clip, width=.47\linewidth]{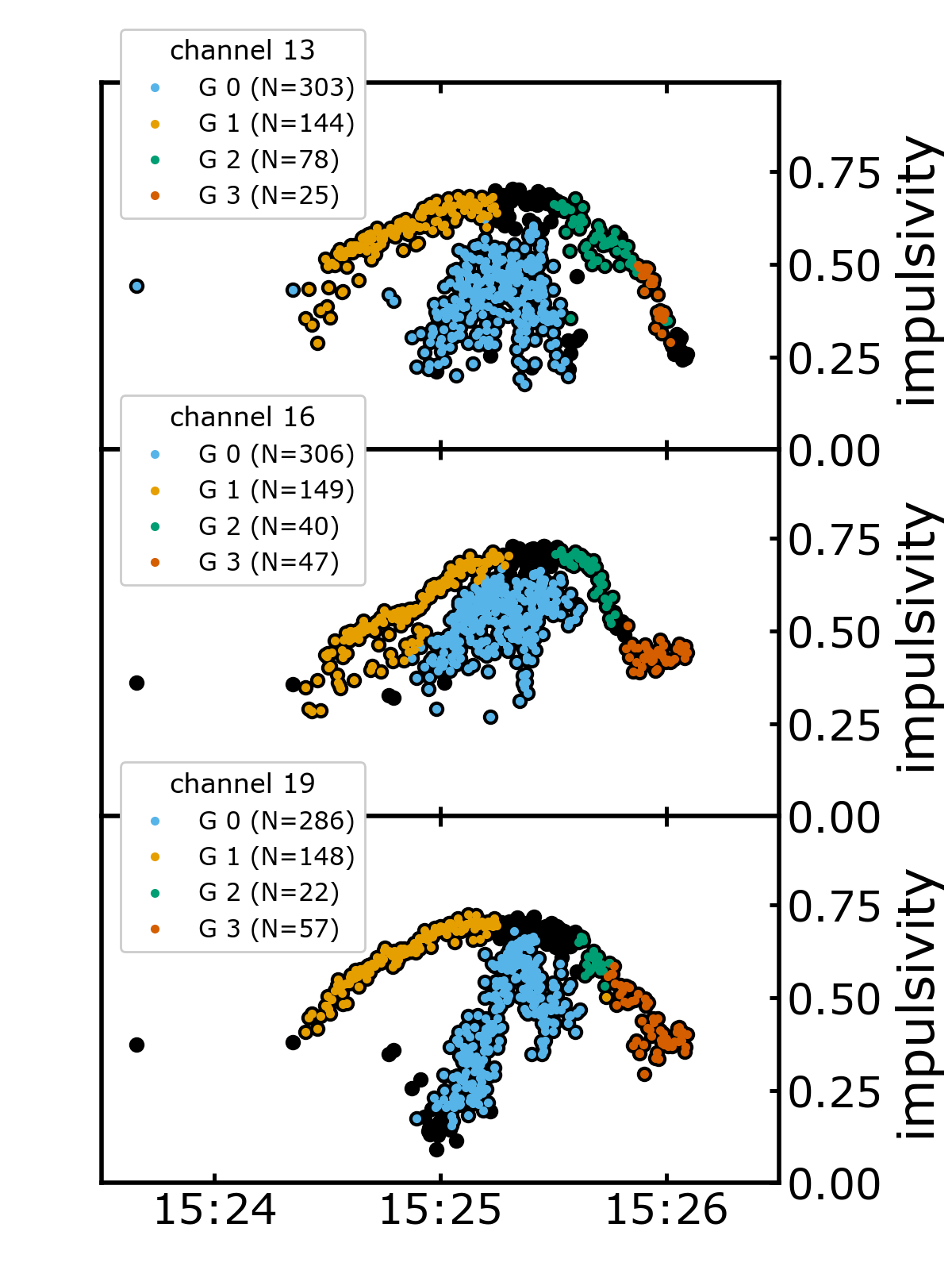}
    \caption{\textit{Left:} Waveforms for two sets of consecutive events seen in channel 16 (upward facing LPDA) from an LC130 on 2022-08-23. The colors match the subsequent cross-correlation-based event grouping. \textit{Right:} Event groups with more than 20 events formed per-channel based on cross-correlation for the three upward facing LPDAs of one station as function of UTC. Events shown in black were not associated to any group. }
    \label{fig:event_classes_xcorr}
\end{figure}

Since planes have a variety of emitters, let alone unintended RFI from e.g.~the turbines, more than one distinct type of signals might be present in the recorded waveforms. This is the case, for example, for an LC130 making a U-turn across the RNO-G array. Looking at a series of events received with the shallow LPDAs, as shown on the left hand side of \autoref{fig:event_classes_xcorr}, one can already by eye identify distinct groups of events. It should be noted that these waveforms are longer than those in \autoref{fig:similar_events_different_stations}. This is predominantly because they are detected with the LPDAs which have a larger group delay than the Vpol antennas and no unfolding has been applied, but a different intrinsic shape is also likely as they are not from the same type of airplane.

In order to group all events originating from the same emitter efficiently, we use pair-wise cross-correlation of all events from the flight. Starting from the pair with highest cross-correlation score $C$ in the set of not-clustered data, we can subsequently add events with high resemblance with at least one other event in the formed group until the remaining events do not surpass a similarity threshold ($C>0.7$ in our case). This approach of greedy cross-correlation grouping allows for a slowly evolving waveform over time (e.g.~$C$ for the first and last event in the group may be smaller than $C=0.7$). One can repeat this procedure to select additional groups of events. Grouping has been done for all three upward-facing antennas for the U-turn flight independently. All three antennas yield comparable groups shown in \autoref{fig:event_classes_xcorr} (right). The largest groups of similar events appear when the aircraft is above the station (G 0), during approach (G 1) and while the plane is leaving (G 2 and 3). Judging from the evolution of impulsivity over time, it seems likely that Groups 1, 2, and 3 are related. However, the waveform is evolving too fast to form a common group with the cross-correlation grouping used. This could be due to a change in the transmitter emission, the sensitivity of the receiving antennas, or both. 

In order to see whether machine learning could improve the grouping, we attempted re-training a classifier developed for wind-induced background events \cite{Laub2024} with aircraft signals but found that the time-difference between the pulses and hence signal arrival direction had a stronger impact on the predicted class than the signal shape. We concluded that perfectly classifying the events is less relevant for a calibration study, where harder cuts can be made and still plenty of signals are available. As such, this did not warrant the development of a dedicated machine learning tool to group the events.

\subsection{Airplanes as a background to neutrino searches}

Neutrino searches typically include selection cuts that reject backgrounds from above the ice surface. However, these cuts will not be 100\% effective and background may be mis-reconstructed to stem from below the surface. Furthermore, in order to increase the effective volume of the detector it is desirable to retain also neutrino vertex locations where the in-ice signal reaches the antennas only after refraction/reflection in the shallow ice. This corresponds to signals arriving locally from above the station, which makes above ground signals a potentially problematic background. 

The simplest and most conservative, but most drastic method of rejecting any potential plane-induced backgrounds would be to exclude data from analysis if an airplane is in the vicinity of the RNO-G array. Vetoing times with an airplane closer than 50\,km leads to an average lifetime loss of 3\%, and may be as high as 12\% for days when there is LC130 traffic to and from Summit station.  Planes with missing position information are too rare to impact the lifetime loss significantly, but they could still be vetoed by setting an appropriate threshold on the Received Signal Strength Indicator (RSSI) of the broadcast ADS-B information. Rejecting all aircraft from which an ADS-B signal is still received would lead to an unacceptable lifetime loss of 60\%. 
The first RNO-G neutrino search will therefore rely on the procedures defined here to identify signals from airplanes as known backgrounds in position and time.

\subsection{Airplanes as calibration source for antenna positions}
\label{sec:position-calibration}

The RNO-G Collaboration is working towards a global antenna positioning fit that will combine information from in-ice pulsers, surface pulsing campaigns, and additional above-surface sources. In the fit, antenna positions (x,y,z) are treated as free parameters. One additional nuisance parameter is the signal travel time in the fiber. Although the fibers were calibrated in the lab prior to deployment, it was found that freezing might affect signal travel times.
The second nuisance parameter to be included in the global fit is the index of refraction profile of the ice (three-part exponential \cite{Windischhofer:2024iK} or fifth order polynomial \cite{Oeyen:2023eN}). To fit a given antenna position, there is thus a total of 11 parameters: 3 from the position, 1 from the fiber time delay, and up to 6 from the ice model. Currently, due to the lack of constraint, only an overall scale for the ice model is allowed to float in the fit; the remaining parameters are fixed. With three arrival time measurements from the local pulsers the system is hence constrained only when the ice profile and the fiber time delay is known, which is not yet the case for RNO-G. Airplane signals will therefore play a crucial role. 
The in-situ pulsers are part of the detector and therefore, in general, cannot constrain an overall rotation of the station and only explore one particular pair of angles. Furthermore, a calibration based on GPS signals cannot be used for antennas deep in the ice.


Given this complexity of the full station calibration procedure, it does not make sense to only use airplane signals for a calibration. We have thus chosen to show here the power of the airplane signals by investigating the positions of the three shallow LPDAs, where it can be assumed that they are at the same depth and ice properties can be ignored. Furthermore, they are connected via coaxial cables and their position is accessible by a differential GPS survey (dGPS) with a resolution of 10 cm, dominated by systematic uncertainties.

\begin{figure}
    \centering
    \includegraphics[width=0.9\linewidth]{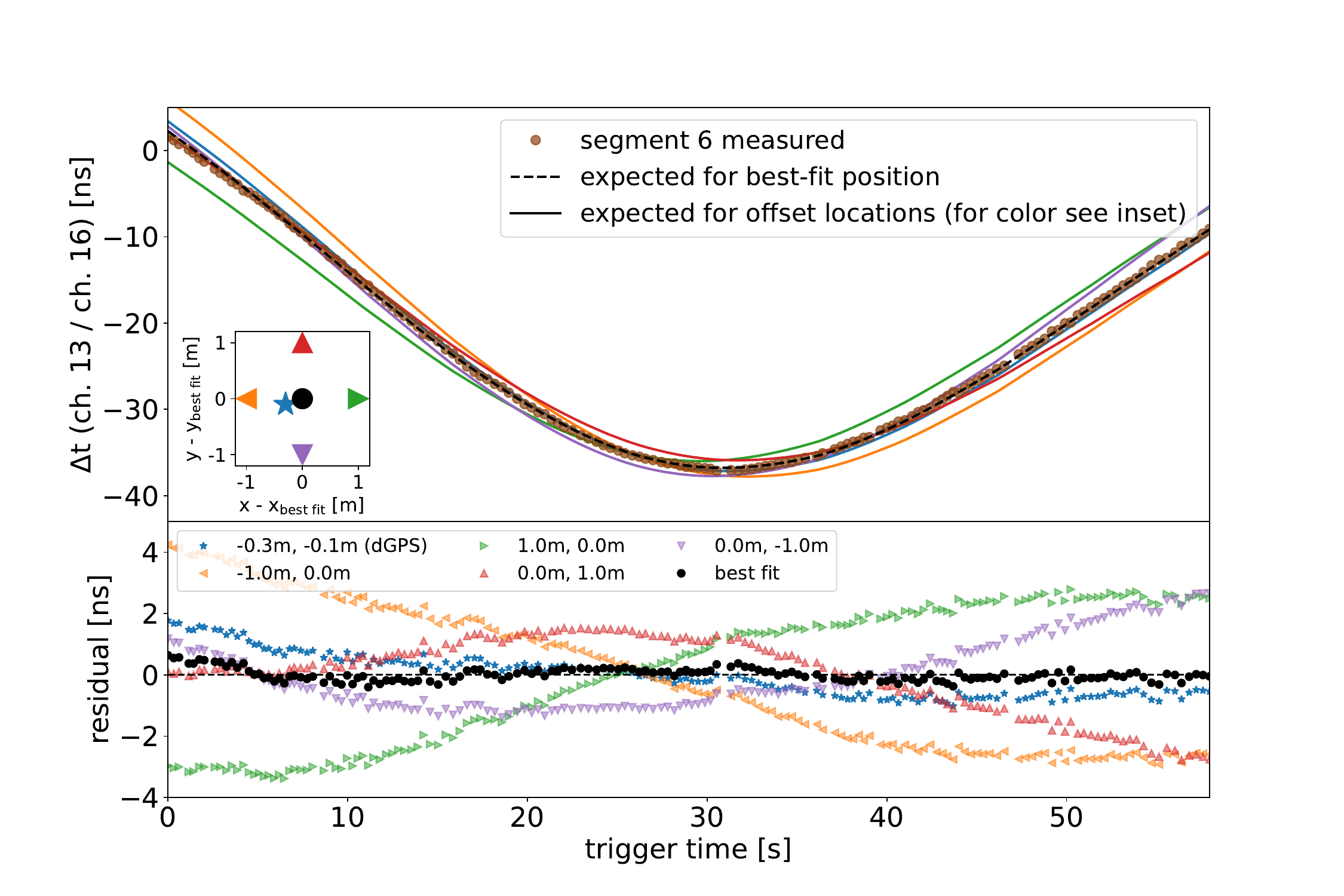}
    \caption{The measured time shift evolution (brown points) compared to the one expected from the aircraft GPS position (black dashed line) for the best-fit position as function of time are shown in the top panel. The shown segment of events corresponds to an LC130 plane flying a half-circle around the station.  Additional lines represent the expectation for the dGPS determined location and four additional points shifted by one meter from the best fit. The bottom panel shows the residual.}
    \label{fig:location_fit}
\end{figure}

If one compares \figref{fig:similar_events_different_stations} and \figref{fig:event_classes_xcorr}, it is clear that the LPDA signals show more features and are longer in contrast to the Vpol signals, which provides an issue in precisely defining the pulse start. Also, due to the arrangement (three different antenna rotations) and the direction dependent antenna pattern, the waveforms for the same signal look different in each antenna, which adds to the complexity of determining the precise time difference between channels. 
We therefore resort to tracking the evolution of the relative time shift of the pulses in the waveforms between channels with time.
This is done as follows: for each antenna, we identify segments (in time) with events belonging to the same event class. For illustration purposes, we first demonstrate the procedure for just one segment in \figref{fig:location_fit}. In each segment, we align all events and store the time shifts needed to align traces.
These relative time shifts between channels are then compared to the expected time shift calculated from the plane position at the time of the events. Due to the uncertainty in pulse start time, we have to leave a constant offset per segment as free parameter, which reduces the precision at which we can calibrate the LPDA positions. In the deep antennas, such a parameter will not be needed, since the signals are short and impulsive see \figref{fig:similar_events_different_stations}, and thus have a clear start time.

Several methods of aligning the waveforms were attempted (e.g.\ cross correlation with an average template, maximum cross correlation out of the last 5 well-matching events) and all yield slowly evolving drifts between the methods of $\mathcal{O}(1\text{ ns})$, which is the systematic uncertainty of this approach. Furthermore, for each segment, a somewhat arbitrary decision needs to be made on the threshold when to still accept traces to be identical, as shown already in \figref{fig:event_classes_xcorr}.

The best fit position using several segments of one flight to cover the full azimuthal range around the station yield best fit positions for channel 16 compared to channel 13 of $\Delta x=0.5 \pm 0.6$ m and  $\Delta y=0.2 \pm 0.8$ m as shift relative to the dGPS measurement. The best fits are thus compatible with the position measured with significant effort during the dGPS survey of the installed station. While the method is less accurate than dGPS (\SI{10}{cm} accuracy), it is much less time-consuming than a survey, can be repeated easily to track detector changes, and may thus be in particular useful for a larger-scale array like IceCube-Gen2 \cite{IceCube-Gen2:2020qha}. For the 12\,m distance between individual surface LPDA antennas, a shift of \SI{80}{cm} can induce a maximum offset of  3.9$^\circ$ for the reconstructed signal arrival direction (0.5$^\circ$ for a 10\,cm shift). 





\section{Multi-station coincidences and absolute station timing}

Another interesting aspect is to use airplane signals detected in coincidence between many stations. In case of neutrino detection, only 10\% of the signals will be recorded by more than one station \cite{RNO-G:2020rmc}. Those \emph{golden} events will be the highest quality neutrino detections, in particular, if the time difference between stations is known. RNO-G does not have a high precision global clock, but each station uses the GPS pulse-per-second (PPS) individually. To check whether the timing would be accurate enough to allow for a combined multi-station reconstruction, rather than two individual reconstructions, signals illuminating several stations from the far-field are exceptionally useful.

\subsection{Selection of multi-station coincidences}

In order to select such coincidence events, we search for clusters of recorded events, where three or more stations simultaneously trigger within less than one microsecond between triggers.  If conservatively assuming the maximum trigger rate of 30 Hz per station, we expect 0.3 such clusters per day for 7 stations caused by simply random coincidences. Requiring in addition the coincidence of such a multi-station cluster with a passing airplane, makes the chance of a random coincidence negligible for the current size of the detector array.

Using multi-station coincidences rejects backgrounds from the ice surface, because it is unlikely that high signal power is refracted into the ice and seen across many stations. Continuous transmission signals and noisy periods typically do not cause isolated strong and impulsive features in the data stream. Since the signal strengths and trigger thresholds vary, individual stations will not trigger on the same feature surpassing the threshold. For example, no such excess in inter-station coincidences was found during solar flares, where elevated signals were present for minutes \cite{Agarwal:2024tat}. 

\begin{figure}
    \centering
    \includegraphics[width=0.5\linewidth]{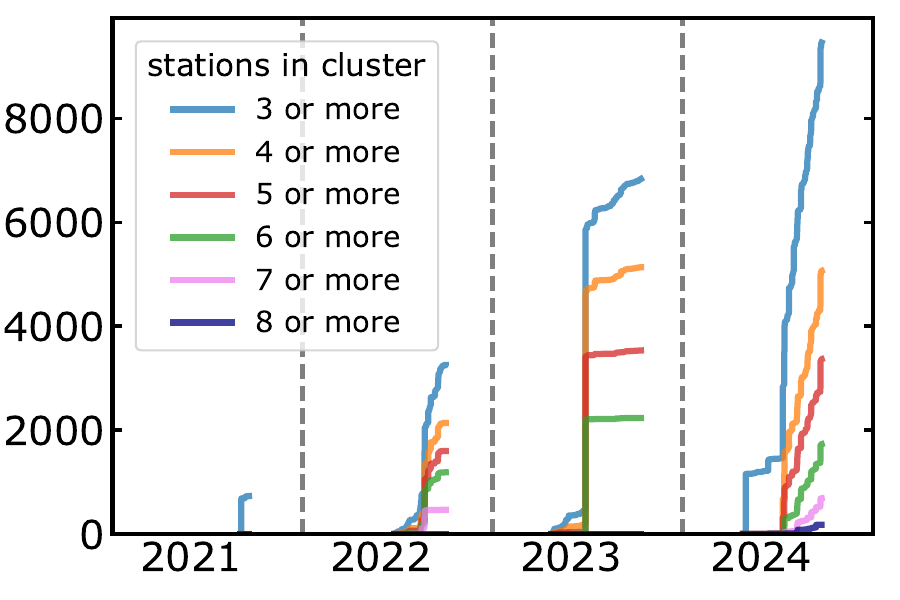}
    \caption{Cumulative number of microsecond-clusters per year observed in RNO-G. Different colors correspond to different numbers of antennas required for the coincidence. Note that in 2021 only three stations were operational and changes to station hardware are typically made during the summer months.}
    \label{fig:clusters_per_year}
\end{figure}

The collected statistics of multi-station coincidences is summarized in \autoref{fig:clusters_per_year} and \autoref{tab:microsecond_cluster_stats}. Since at least three stations must be operational, the criterion is only met during the summer months. Nonetheless, a significant data-set is available for to evaluate the detector performance. 

\begin{table}
    \centering
    \caption{Events recorded in RNO-G in clusters of at least 3 stations and less than a microsecond between station trigger times.}
    \label{tab:microsecond_cluster_stats}
    \begin{tabular}{|c|r|r|r|r|}
    \hline
year & clusters & triggered events & deep trigger& shallow trigger \\ 
 &  ($\geq$ 3 stations) &  (all stations) &  & \\ 
\hline
2021 & 730 & 2191 & 1458 & 736 \\
2022 & 3255 & 15165 & 7393 & 7743\\
2023 & 6836 & 31434 & 20041 & 11390\\
2024 & 9464 & 39435 & 8053 & 31362\\
\hline
\end{tabular}
\end{table}

\begin{figure}
    \centering
\includegraphics[width=0.55\linewidth, trim=3cm 0 3cm 0, clip]{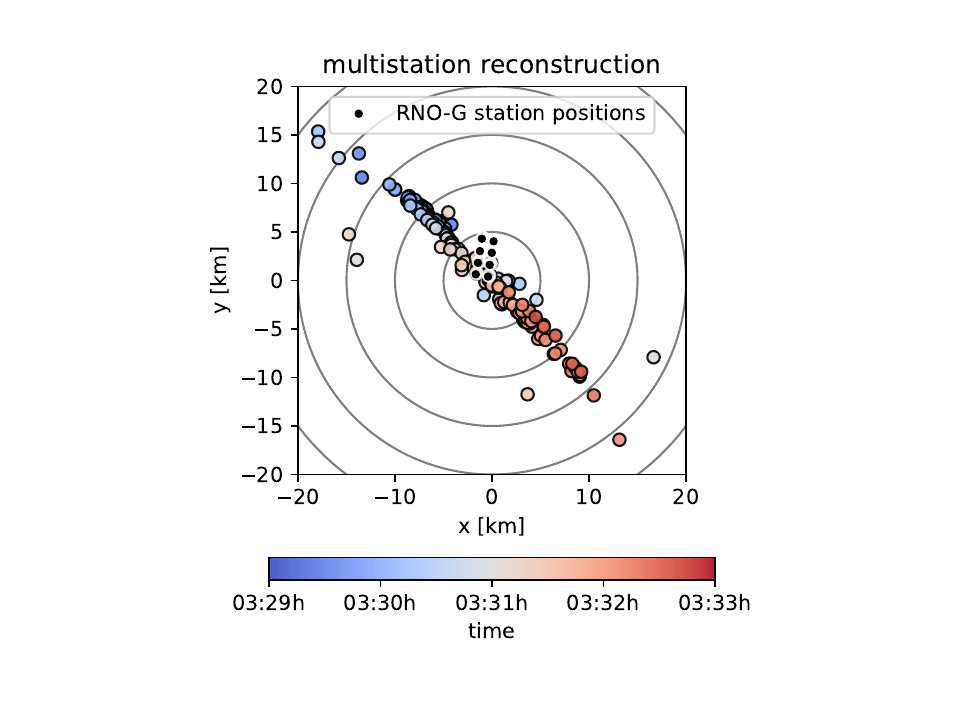}
    \hfill
\includegraphics[width=0.44\linewidth]{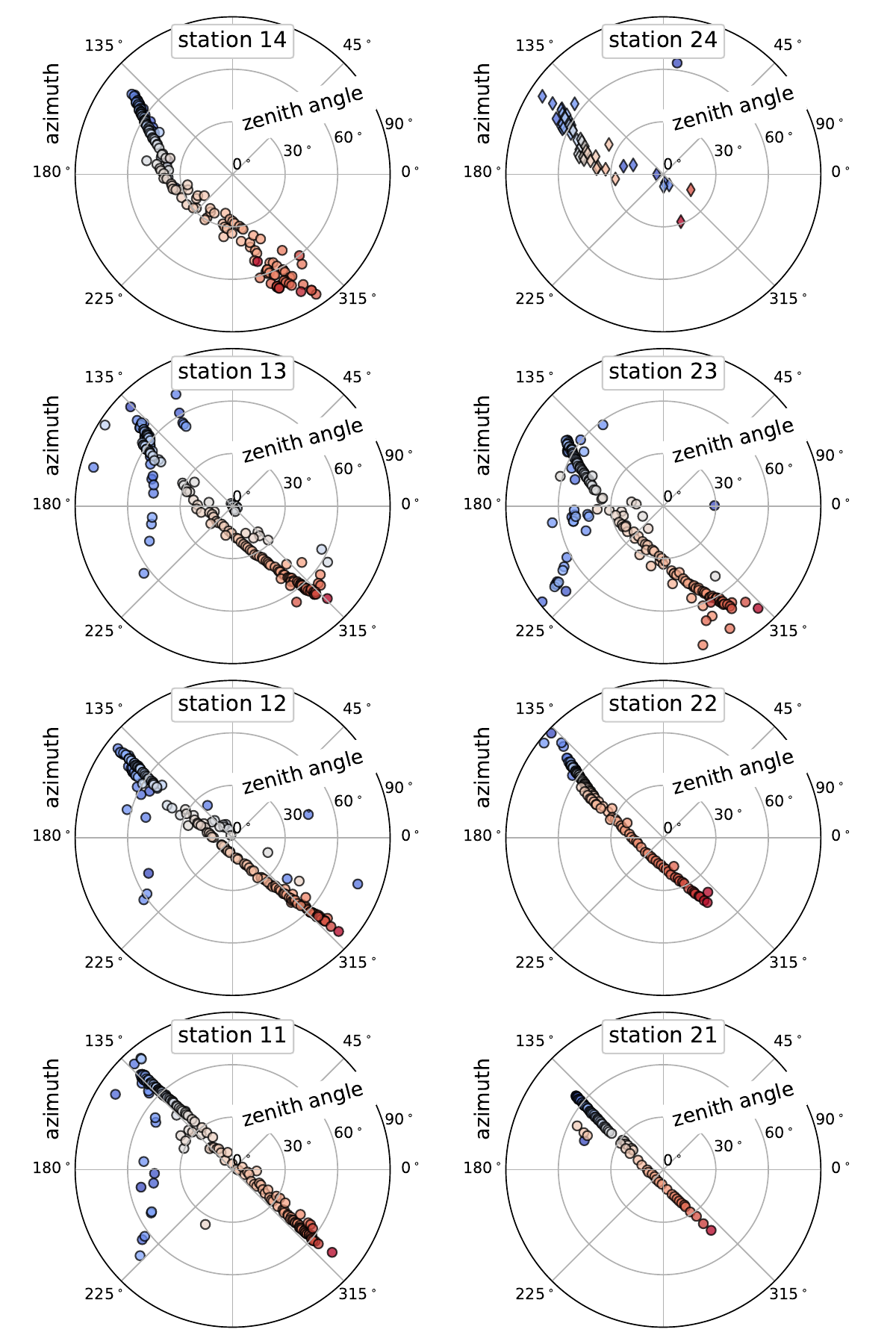}
    \caption{Reconstructed aircraft position for a C17 flight without transmitted position. Time is color-coded. \textit{Left}: Reconstruction based on the trigger times from the stations for multi-station coincident events. \textit{Right:} Reconstruction of zenith and azimuth angles for individual stations based on the signal waveforms for shallow (circles) and deep (diamonds) triggered events. No quality cuts on either reconstruction have been applied.}
    \label{fig:C17_reconstruction}

\end{figure}

In 2024, new DAQ-Boxes were installed in all but one station (labeled as station 24). One of the changes in readout electronics was lowering the threshold for surface triggers, which leads to an increase in multi-station coincidences and more shallow triggers with respect to deep triggers. If at least three stations trigger coincidentally, the trigger timestamps can be used to triangulate the point of emission. For a US military C17 aircraft that did not send position info via ADS-B, the reconstructed positions using multi-station coincident timing are shown in \autoref{fig:C17_reconstruction} (left). A plane-wave fit is able to reconstruct the arrival direction of the signal using the signal delay in the shallow LPDAs, or the Vpols below -80\,m depth, respectively, albeit with a very coarse resolution. The heading of the track is consistent for all eight currently installed detector stations, with a precise reconstruction still depending on individual station calibration. Owed to the higher trigger threshold for the surface component for station 24 with the old DAQ system, almost all coincidence triggers for this station were issued by the deep phased array. While the aircraft did not send GPS positions, the approach is noticeable in the data by an increased RSSI for the received ADS-B messages. Multi-station coincidences reconstruct an altitude of 9.8\,km above sea-level and a velocity of 830\,km/h, which is consistent with the cruise altitude and speed of these types of airplanes. While a straight travel track can be fitted for the aircraft in \autoref{fig:C17_reconstruction}, RNO-G is only planning to use flights that do broadcast their position via ADS-B for calibration.

Of interest for further analysis are those multi-station events that are not associated with an airplane. In principle, they could stem from cosmic ray air showers or other far-field sources that constitute a background. However, defining a clear non-association with airplanes is hard to generalize and special cases tend to appear. Reconstructing multi-station events ($> 2$ stations) where no valid ADS-B message was received in coincidence (RSSI $< -30$), still shows a strong preference in arrival direction for \emph{typical} aircraft routes. This means that those coincidences are likely due to airplanes, but the cuts are not efficient. With more strict assumptions ($> 3$ stations, RSSI $< -25$), we retain 197 events that will be subject to further investigation. Initial results suggest that a large fraction reconstructs towards the direction of Summit Station, linking them to local human activity.

\subsection{Absolute timing between stations}

\begin{figure}
    \centering
    \includegraphics[width=0.5\linewidth]{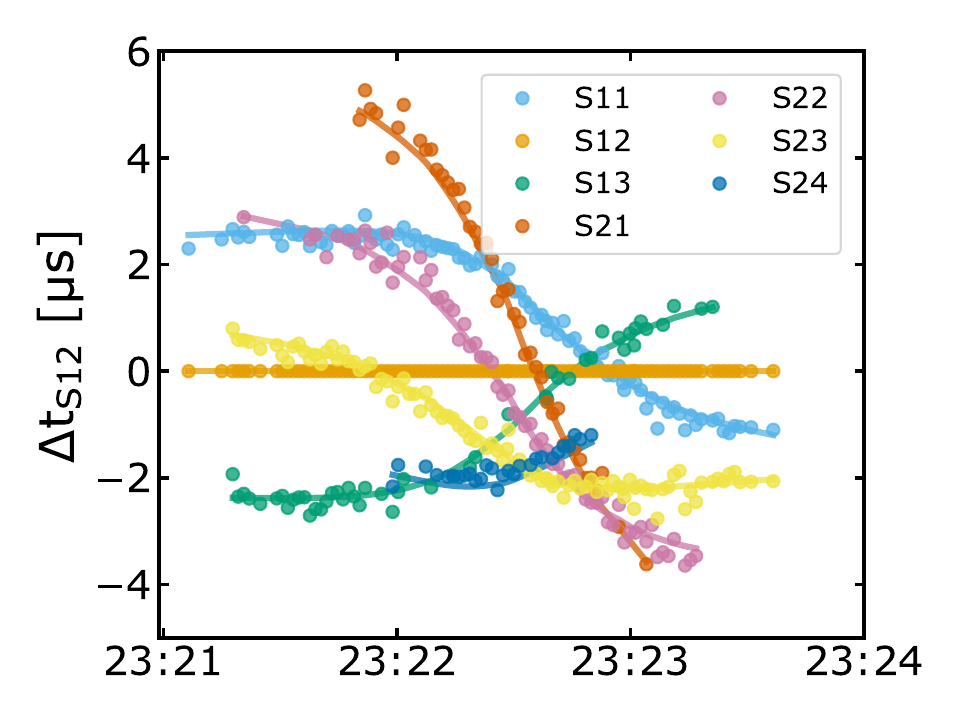}
    \includegraphics[width=.49\linewidth]{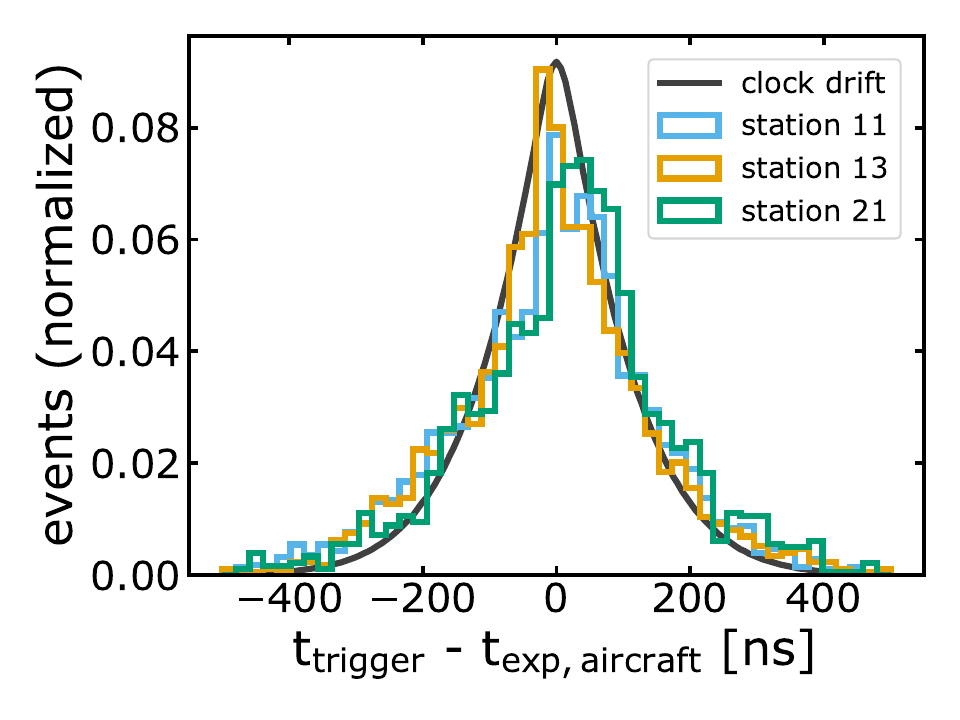}
    \caption{\textit{Left:} Trigger time differences ($\Delta t_{S12}$) of all stations relative to station 12 (S12) for a B777 flying from Europe to Asia. The line indicates the expectation from the position of the aircraft. Right: Measured difference between expected and actual signal arrival time using events triggering the shallow component. For clarity only three stations are shown, the other four stations operational at the time perform similarly.  Also shown is the expected distribution based on oscillator drift alone (black line).}
    \label{fig:time-difference}
\end{figure}

We can use coincident events to determine the inter-station timing accuracy of RNO-G.
For this, we compare the difference in the trigger time between individual stations with the expected difference of signal arrival times using the aircraft position at the time when the event was recorded.  Since the time of emission is unknown, all times are taken relative to one reference station. This procedure is illustrated in \autoref{fig:time-difference} (left) taking station 12 as reference. 

Taking a distribution of expected and measured time differences (\autoref{fig:time-difference}, right) shows a comparable spread of $\SI{270}{ns}$ (68\% range) and $\SI{490}{ns}$ (95\% range) for different stations. The uncertainty of the absolute timing in the RNO-G stations is dominated the stability of the 100\,MHz clock determining the subsecond part of the trigger time. This clock consists of a counter driven by a local oscillator with a frequency close to 100\,MHz counting clicks $K$ at each PPS from GPS and whenever a trigger is issued. The subsecond part is then calculated as \begin{equation}
    t_\mathrm{sub-sec} = \frac{K(\mathrm{trigger}) - K(\mathrm{last\ PPS})}{K(\mathrm{last\ PPS}) - K(\mathrm{last\ to\ last\ PPS})}.
    \label{eq:sysclk_time}
\end{equation}
Using the recorded $K(\mathrm{last\ PPS})$ and $K(\mathrm{last\ to\ last\ PPS})$ we find, that the spread of counts between seconds is 14.1 clicks and similar for all stations. For a 100\,MHz oscillator folded with a uniform distribution of the subsecond part the resulting uncertainty shown as 'clock drift' in \autoref{fig:time-difference} (right). The distribution matches the obtained spread found in data. We hence conclude that the stability of the clock is the main driver for the uncertainty and that interferometric reconstructions with multiple station that requires sub-nanosecond scale timing is not feasible with the current set-up. 

Note that the linearly growing drift for larger sub-second parts could be avoided by taking the next and last PPS count in the denominator in \autoref{eq:sysclk_time}. Since the next PPS count is not known at the time of writing an event, this number could be obtained from the recorded data using the following events, and is only available provided the trigger rates are high enough. This procedure is not only cumbersome, but would also not avoid the general problem of a drifting clock speed within a second.

In addition to this dominant uncertainty from the local oscillator, which induces a spread of about \SI{270}{ns} in the absolute timing of RNO-G stations, the other identified uncertainties are small. They conservatively add up to an additional 50\,ns smearing in trigger time. These are (1) the absolute precision of the PPS: 10\,ns, (2) the uncertainty from the time between subsequent clicks: 10\,ns for 100\,MHz, (3) it is currently not recorded which two channels fired the trigger where on the pulse, leading to: $\sim$30\,ns, and (4) the uncertainty from the block-wise digitizer readout after a trigger has been issued: 40\,ns.

While items (1) and (2) are (almost) irreducible uncertainties, item (3) could be avoided by changing the firmware or by matching up waveforms between different stations. Uncertainty (4) is non-trivial to eliminate in the current DAQ setup. When a trigger is issued, the recorded waveforms are read from the digitizer buffer \cite{Roberts:2018xyf} in blocks of 128 samples, and the offset between the time of the trigger with respect to the start of the next block is not recorded.

\section{Other opportunities for calibration}

The main use case of the airplane signals will be to calibrate the position of the antennas, in combination with mapping out the index of refraction profile, as both are entangled. To calibrate the position of an antenna in the ice, only the timing of the received pulses is used.
However, in addition to the uncertainties of the antenna positions, in particular the azimuth orientation of the LPDA antennas, which are placed in hand-dug trenches is hard to determine better than 10--20$^\circ$ during deployment. It is then further challenging to monitor whether the LPDAs remain aligned when snow accumulates and the set-up sinks below the surface. In \cite{Beise:2022stx} an in-situ calibration system was evaluated, which is, however, mostly sensitive to snow accumulation rather than antenna orientation. 

\begin{figure}
    \centering
    \includegraphics[width=0.9\linewidth]{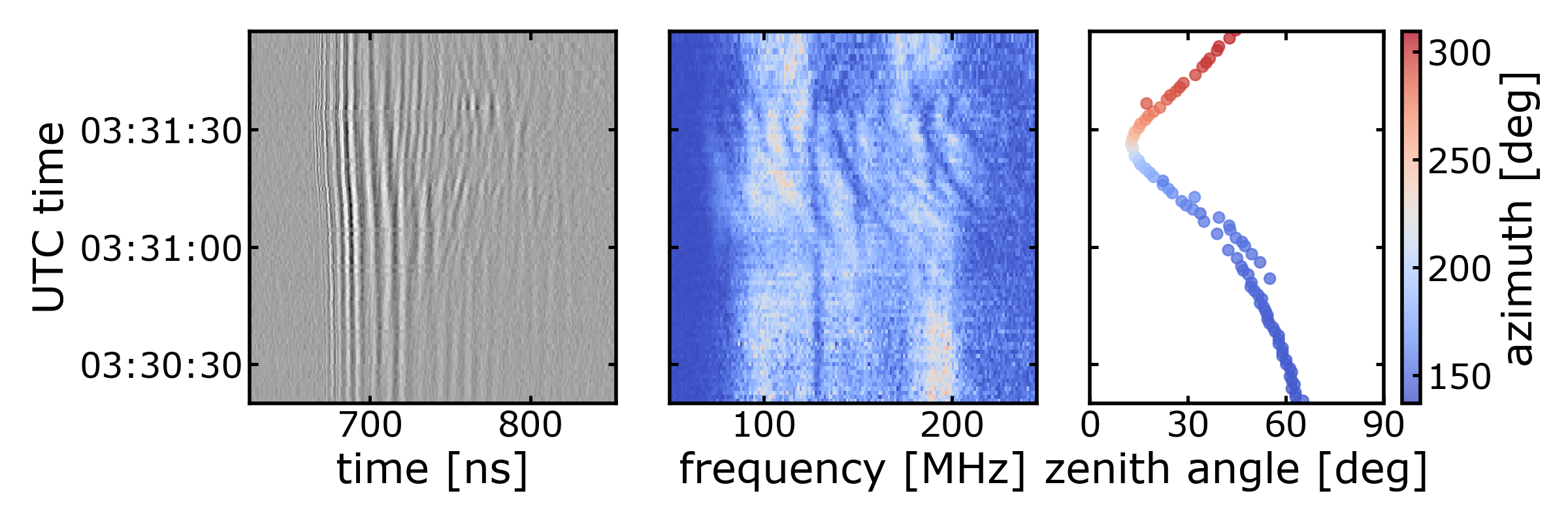}
    \caption{Structures in time-domain (left, aligned along the high frequency beginning of the waveform at $\sim$680\,ns) and in the frequency spectra (center, blue to white) observed in the shallow LPDA channels (here station 22, channel 13) during overhead passage of an aircraft on 08/08/2024. The right panel shows the reconstructed signal arrival direction from a plane-wave fit. }
    \label{fig:zenith_dependent_structures}
\end{figure}

Impulsive events from airplanes might be useful in calibrating the antenna orientation by tracking and comparing the signal strength of different antennas over time. Qualitatively, the effect of antenna orientation and reflection at the ice surface was evident in the arcs of mis-reconstructed events in \autoref{fig:C17_reconstruction} (right), where the antennas with tines oriented orthogonal to the aircraft track did not see a strong pulse at large distances. With only two pulses, this led to an under-constrained fit of the direction that appear as arcs in the plot.

\autoref{fig:zenith_dependent_structures} shows the waveforms observed in one LPDA aligned by matching the high-pass filtered traces via cross-correlation, as it was also used for the position calibration in Section \ref{sec:position-calibration} to obtain time differences. The figure also shows the corresponding frequency spectrum and the reconstructed zenith angle of the aircraft. Assuming that the emitted pulse-shape is the same as a function of time, it is evident that the received pulse changes as a function of zenith angle as the aircraft passes by the station. Other LPDA channels in the same and other stations show a comparable evolution in time. The results of the simple plane-wave fit and the second arch illustrate that the changing pulse shape (as a function of arrival direction) has to be accounted for to avoid mis-reconstructions. We attribute this slow change of the observed pulse shape to the change in sensitivity of the antenna, which means that the correct modeling of the antenna pattern should be possible using this data. 
It is known that the antenna pattern of the receiving LPDAs changes as a function of orientation, which may allow to determine the accurate positioning or at least constrain the modeling of the antenna response. 
A complication when constraining the antenna pattern of the RNO-G LPDAs in-situ is the unknown characteristics of the emitter, which puts the study of the antenna orientation beyond the scope of this paper. 


\section{Conclusion and outlook}
As we have shown in this paper, airplanes emit various impulsive radio signals along their flight paths that are recorded with RNO-G. On their own, the often highly impulsive signals constitute a background to neutrino searches with RNO-G. By recording the transmitted airplane positions at Summit Station, it is ensured that these backgrounds can be properly vetoed. 

The critical impulsiveness of the signals, however, also makes them a very suitable calibration signal for radio neutrino arrays. We have shown that using airplane signals alone the positions of the shallow LPDAs can be calibrated to a precision of half a meter, almost rivaling differential GPS surveys with no additional field work needed. The signals selected in this analysis will be used in a forthcoming comprehensive instrument calibration for RNO-G that solves for the precise antenna positions, residual instrumental delays, and the ice profile at the same time. Aircraft signals will also allow us to repeat the calibration each year and monitor the instrument performance over time. The presented data-set already allowed us to probe the realized timing resolution of RNO-G, which shows a spread of \SI{270}{ns}, which matches expectations due to the drift of the local oscillator. 

\begin{figure}
    \centering
    \includegraphics[width=0.5\linewidth]{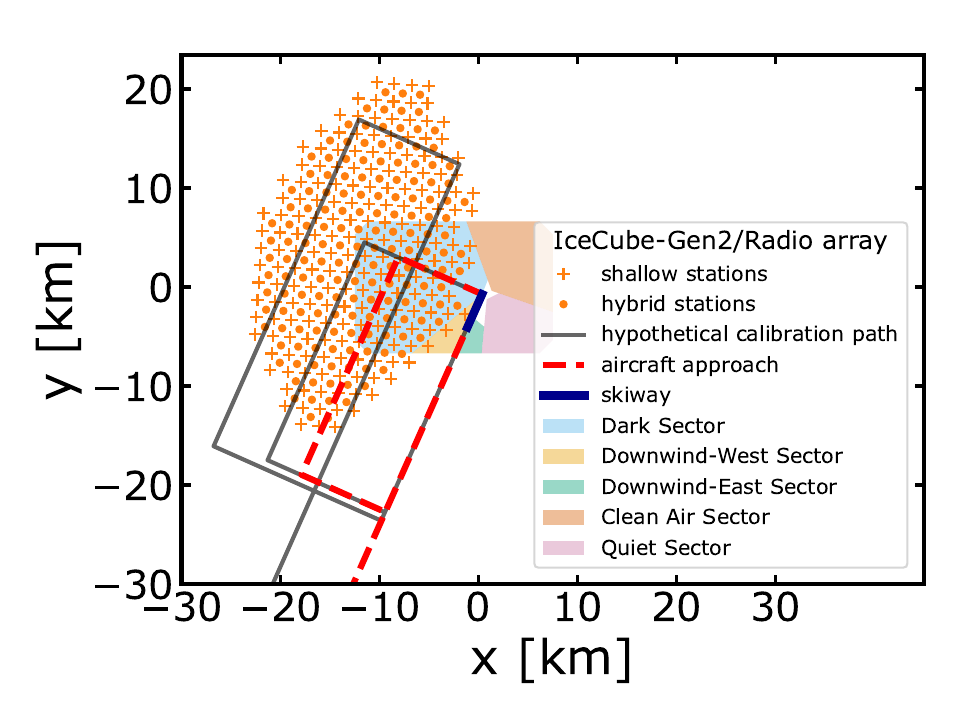}
    \caption{Map of the planned radio array of IceCube-Gen2 \cite{IceCube-Gen2:2020qha} and the sectors at the South Pole. The nominal approach path for landing airplanes crosses only part of the planned array. Additional loops would require additional $\mathcal{O}$(125\,km) of flight distance as indicated on the map. Such a loop would significantly improve coverage, naturally scaling with the amount of different azimuths covered per station. }
    \label{fig:sout_pole_map}
\end{figure}

Beyond RNO-G, there are plans for an even larger radio neutrino array at the South Pole as part of IceCube-Gen2 \cite{IceCube-Gen2:2020qha}. The calibration effort for such an instrument scales with size, if relying on drones or surveying campaigns. Also, using solar flares for calibration (see \cite{Agarwal:2024tat}) will be challenging due to the low angle of the Sun above the horizon at the South Pole.
Airplanes provide a useful alternative here. Unfortunately, there are no commercial plane routes close to South Pole that one could use the emission of, as it is possible in Greenland. However, similarly to Summit Station, during the Antarctic Summer, LC130s  (and other smaller planes like Baslers or Twin Otters) fly in and out of the Amundsen--Scott South Pole station. As shown in \autoref{fig:sout_pole_map}, the observed approach routes towards station, happen on the same length scales as the extent of the radio array of IceCube-Gen2. One can envision to plan an additional circle across the array for calibration purposes, which will use limited additional resources and will provide very useful calibration data points, removing the need for extensive surveys. The figure also serves as a reminder that the distance-scales of IceCube-Gen2 are larger than easily reached with small dedicated calibration drones.

\section*{Acknowledgments}

We are thankful to the support staff at Summit Station for making RNO-G possible. We also acknowledge our colleagues from the British Antarctic Survey for building and operating the BigRAID drill for our project.

We would like to acknowledge our home institutions and funding agencies for supporting the RNO-G work; in particular the Belgian Funds for Scientific Research (FRS-FNRS and FWO) and the FWO programme for International Research Infrastructure (IRI), the National Science Foundation (NSF Award IDs 2112352, 2111232, 2111410, 2411590, and collaborative awards 2310122 through 2310129), and the IceCube EPSCoR Initiative (Award ID 2019597), the Helmholtz Association, the Swedish Research Council (VR, Grant 2021-05449 and 2021-00158), the University of Chicago Research Computing Center, and the European Union under the European Unions Horizon 2020 research and innovation programme (grant agreements No 805486), as well as (ERC, Pro-RNO-G No 101115122 and NuRadioOpt No 101116890).

\bibliographystyle{JHEP}
\bibliography{Bibliography}

\end{document}

%% file: authorlist.tex
\author[1]{S.~Agarwal}
\author[2]{J.~A.~Aguilar}
\author[3]{N.~Alden}
\author[1]{S.~Ali}
\author[4]{P.~Allison}
\author[5]{M.~Betts}
\author[1]{D.~Besson}
\author[6]{A.~Bishop}
\author[7]{O.~Botner}
\author[8]{S.~Bouma}
\author[9,10]{S.~Buitink}
\author[2]{R.~Camphyn}
\author[6]{J.~Chan}
\author[2]{S.~Chiche}
\author[11]{B.~A.~Clark}
\author[7]{A.~Coleman}
\author[1]{K.~Couberly}
\author[12]{S.~de~Kockere}
\author[12]{K.~D.~de~Vries}
\author[3]{C.~Deaconu}
\author[13]{P.~Giri}
\author[7]{C.~Glaser}
\author[7]{T.~Gl{\"u}senkamp}
\author[4]{H.~Gui}
\author[7]{A.~Hallgren}
\author[8,14]{S.~Hallmann}
\author[15]{J.~C.~Hanson}
\author[16]{K.~Helbing}
\author[5]{B.~Hendricks}
\author[14,8]{J.~Henrichs}
\author[7]{N.~Heyer}
\author[1]{C.~Hornhuber}
\author[14]{E.~Huesca Santiago}
\author[4]{K.~Hughes}
\author[14,8]{A.~Jaitly}
\author[14]{T.~Karg}
\author[6]{A.~Karle}
\author[6]{J.~L.~Kelley}
\author[3]{J.~Kimo}
\author[8]{C.~Kopper}
\author[2,12]{M.~Korntheuer}
\author[14,17]{M.~Kowalski}
\author[13]{I.~Kravchenko}
\author[5]{R.~Krebs}
\author[6]{M.~Kugelmeier}
\author[8]{R.~Lahmann}
\author[13]{C.-H. Liu}
\author[18]{M.~J.~Marsee}
\author[14,8]{Z.~S.~Meyers}
\author[10]{K.~Mulrey}
\author[6]{M.~Muzio}
\author[14,8]{A.~Nelles}
\author[19]{A.~Novikov}
\author[4]{A.~Nozdrina}
\author[3]{E.~Oberla}
\author[20]{B.~Oeyen}
\author[19]{N.~Punsuebsay}
\author[14,20]{L.~Pyras}
\author[7]{M.~Ravn}
\author[16]{A.~Rifaie}
\author[20]{D.~Ryckbosch}
\author[8]{O.~Schlemper}
\author[2]{F.~Schl{\"u}ter}
\author[9,22]{O.~Scholten}
\author[19]{D.~Seckel}
\author[1]{M.~F.~H.~Seikh}
\author[20]{J.~Stachurska}
\author[12]{J.~Stoffels}
\author[2]{S.~Toscano}
\author[6]{D.~Tosi}
\author[5]{J.~Tutt}
\author[9,12]{D.~J.~Van Den Broeck}
\author[12]{N.~van Eijndhoven}
\author[3]{A.~G.~Vieregg}
\author[11]{A.~Vijai}
\author[3]{C.~Welling}
\author[18]{D.~R.~Williams}
\author[3]{P.~Windischhofer}
\author[5]{S.~Wissel}
\author[1]{R.~Young}
\author[8]{A.~Zink}
\affiliation[1]{University of Kansas, Dept.\ of Physics and Astronomy, Lawrence, KS 66045, USA }
\affiliation[2]{Universit\'e Libre de Bruxelles, Science Faculty CP230, B-1050 Brussels, Belgium}
\affiliation[3]{Dept.\ of Physics, Enrico Fermi Inst., Kavli Inst.\ for Cosmological Physics, University of Chicago, Chicago, IL 60637, USA }
\affiliation[4]{Dept.\ of Physics, Center for Cosmology and AstroParticle Physics, Ohio State University, Columbus, OH 43210, USA }
\affiliation[5]{Dept.\ of Physics, Dept.\ of Astronomy \& Astrophysics, Center for Multimessenger Astrophysics, Institute of Gravitation and the Cosmos, Pennsylvania State University, University Park, PA 16802, USA }
\affiliation[6]{Wisconsin IceCube Particle Astrophysics Center (WIPAC) and Dept.\ of Physics, University of Wisconsin-Madison, Madison, WI 53703,  USA }
\affiliation[7]{Uppsala University, Dept.\ of Physics and Astronomy, Uppsala, SE-752 37, Sweden }
\affiliation[8]{ Erlangen Centre for Astroparticle Physics (ECAP), Friedrich-Alexander-University Erlangen-N\"urnberg, 91058 Erlangen, Germany }
\affiliation[9]{Vrije Universiteit Brussel, Astrophysical Institute, Pleinlaan 2, 1050 Brussels, Belgium }
\affiliation[10]{Dept.\ of Astrophysics/IMAPP, Radboud University, PO Box 9010, 6500 GL, The Netherlands }
\affiliation[11]{ Department of Physics, University of Maryland, College Park, MD 20742, USA }
\affiliation[12]{Vrije Universiteit Brussel, Dienst ELEM, B-1050 Brussels, Belgium }
\affiliation[13]{Dept.\ of Physics and Astronomy, Univ.\ of Nebraska-Lincoln, NE, 68588, USA }
\affiliation[14]{ Deutsches Elektronen-Synchrotron DESY, Platanenallee 6, 15738 Zeuthen, Germany }
\affiliation[15]{Whittier College, Whittier, CA 90602, USA }
\affiliation[16]{Dept. of Physics, University of Wuppertal, D-42119 Wuppertal, Germany}
\affiliation[16]{ Institut f\"ur Physik, Humboldt-Universit\"at zu Berlin, 12489 Berlin, Germany }
\affiliation[17]{ Dept.\ of Physics and Astronomy, University of Alabama, Tuscaloosa, AL 35487, USA }
\affiliation[18]{ Dept.\ of Physics and Astronomy, University of Delaware, Newark, DE 19716, USA }
\affiliation[19]{ Ghent University, Dept.\ of Physics and Astronomy, B-9000 Gent, Belgium }
\affiliation[20]{Department of Physics and Astronomy, University of Utah, Salt Lake City, UT 84112, USA}
\affiliation[21]{Kapteyn Institute, University of Groningen, PO Box 800, 9700 AV, The Netherlands }